\documentclass[aps,prb,twocolumn,amsmath,amssymb,superscriptaddress,floatfix]{revtex4-2}

\usepackage{amsmath}
\usepackage{xcolor}
\usepackage{graphicx}% Include figure files
\usepackage{dcolumn}% Align table columns on decimal point
\usepackage{bm}% bold math
\usepackage{verse}
\usepackage{here}
\usepackage{graphicx}%
\usepackage{bm}%
\usepackage{graphicx,bm}
\usepackage{pifont}
\usepackage{amsmath,amssymb}
\usepackage{color}

\begin{document}

\title{Doping control of magnetism and emergent electromagnetic induction in high-temperature helimagnets}

\author{Aki Kitaori$^{1}$, Jonathan S. White$^{2}$, Naoya Kanazawa$^{1}$, Victor Ukleev$^{2}$, Deepak Singh$^{2}$, Yuki Furukawa$^{1}$, Taka-hisa Arima$^{3}$, Naoto Nagaosa$^{1,4}$ and Yoshinori Tokura$^{1,4,5}$}
\affiliation{$^{1}$ Department of Applied Physics, University of Tokyo, Tokyo 113-8656, Japan \\ 
$^{2}$ Laboratory for Neutron Scattering and Imaging (LNS), Paul Scherrer Institute (PSI), CH-5232 Villigen, Switzerland \\
$^{3}$ Department of Advanced Materials Science, University of Tokyo, Kashiwa 277-8561, Japan \\
$^{4}$ RIKEN Center for Emergent Matter Science (CEMS), Wako 351-0198, Japan \\
$^{5}$ Tokyo College, University of Tokyo, Tokyo 113-8656, Japan
}

\date{\today}

\begin{abstract}
Ac current-driven motions of spiral spin textures can give rise to emergent electric fields acting on conduction electrons. This in turn leads to the emergent electromagnetic induction effect which may realize quantum inductor elements of micrometer size. ${\rm YMn}_{6}{\rm Sn}_{6}$ is a helimagnet with a short helical period (2-3 nm) that shows this type of emergent inductance beyond room temperature. To identify the optimized materials conditions for ${\rm YMn}_{6}{\rm Sn}_{6}$-type room-temperature emergent inductors, we have investigated emergent electromagnetic inductance (EEMI) as the magnetism is modified through systematic partial substitution of Y by Tb. By small angle neutron scattering and inductance measurements, we have revealed that the pinning effect on the spin-helix translational mode by Tb doping selectively and largely suppresses the negative component of EEMI, while sustaining the positive inductance arising from the spin tilting mode. We also find that in addition to the spin helix, even the spin-collinear antiferromagnetic structure can host the positive EEMI due to thermally enhanced spin fluctuations. The present study highlights the facile control of both the magnitude and sign of EEMI beyond room temperature, and thus suggests a route to expand the range of emergent inductor candidate materials. 

\end{abstract}

\maketitle

\section{I. Introduction}
The inductor, a most important element of contemporary electric circuits, is characterized by the relation $V$ = $L dI/dt$, where $V$, $I$, and $L$ are voltage, current and inductance, respectively. Since $L$ of the conventional inductor coil is in proportion to ${n}^{2}{A}$, $n$ being the number of coil windings, and $A$ the coil cross-section, it is technically difficult to reduce the dimension of the inductor of coil form. Recently, a new and simpler scheme of the electromagnetic induction has been proposed which may dramatically miniaturize the inductor element, namely via the use of current-induced spin dynamics in a helical-spin system~\cite{fref1}. The idea is based on the time-dependent emergent electromagnetic field, or Berry phase dynamics~\cite{fref2, fref3} of the conduction electrons flowing on a helical spin texture~\cite{fref1}.\\
\ Noncoplanar spin textures, as exemplified by magnetic skyrmions and spin hedgehogs~\cite{fref4}, are endowed with a non-zero scalar spin chirality $\bm{S}_i \cdot (\bm{S}_j \times \bm{S}_k)$: $i$, $j$ and $k$ being neighboring spin sites, that exerts emergent magnetic field on conduction electrons~\cite{fref2, fref3}. The effect of scalar spin chirality has been investigated via topological or geometrical Hall effects stemming from the real-space emergent magnetic field ${\bm b}$ ~\cite{fref5, fref6, fref7}; here ${b}_{i} = \frac{\hbar}{2e}{\epsilon}_{ijk}{\bm n}\cdot ({\partial}_{j}{\bm n}\times {\partial}_{k}{\bm n})$, ${\epsilon}_{ijk}$ is the Levi-Civita symbol and ${\bm n}={\bm S}/|{\bm S}|$. In recent years, the dynamics of emergent fields have also attracted great attention ~\cite{fref1, fref8}. The time ($t$) evolution of emergent magnetic field (${\bm b}$) can give rise to the emergent electric field (${\bm e}$) according to the generalized Faraday's law~\cite{fref9}, $\nabla \times {\bm e}= -d{\bm b}/dt$, or equivalently ${e}_{i} = \frac{\hbar}{e}{\bm n}\cdot ({\partial}_{i}{\bm n}\times {\partial}_{t}{\bm n})$, giving rise to the emergent electromagnetic induction phenomenon~\cite{fref9, fref10, fref11, fref12, fref13, fref14, fref15}. Among the ideas to detect or utilize the emergent electric field, the emergent electromagnetic inductance (EEMI) based on dynamical spin spiral structures stands out in light of the large signal magnitude~\cite{fref1}. Spiral spin textures are noncollinear with the vector spin chirality $\bm{S}_i \times \bm{S}_j$, but occasionally take coplanar structure (e.g., proper screw) with no static scalar spin chirality $\bm{S}_i \cdot (\bm{S}_j \times \bm{S}_k)$. Nevertheless, the dynamically-swept noncoplanar spin structure under an ac current excitation along the spin helix propagation vector (${\bm q}$) can generate the emergent electric field ${\bm e}$ along the direction of ${\bm q}$, as expressed as ${e}_{q}=\frac{\hbar q}{2e}\frac{\partial m}{\partial t}$, $m$ being the projection of ${\bm n}$ along ${\bm q}$, according to the above-described formula of ${\bm e}$~\cite{fref1}. This emergent electric field becomes 90\(^\circ\) out of phase with the applied ac current, and can be equated with the electric field caused by the imaginary part of the complex impedance. The intensity of emergent electric field gets larger when the ac current frequency is higher, due to the time-derivative term of spin motion. Therefore, this imaginary impedance behaves as the inductance $L$. Here, note that $L$ is expected to be proportional to ${\bm q}$; namely the shorter helix period ${\it \lambda}$ is favorable to attain the larger $L$. This is the emergent electromagnetic inductance (EEMI) caused by the spin-tilting mode out of the spin spiral plane, as shown in Fig.~1(a). To be precise, the ac current excitation on the spin helix can induce not only the tilting mode but also the spin-helix translational mode or phason mode [Fig.~1(b)], in which the helical spin rotates uniformly within the plane and hence appears to propagate along the ${\bm q}$ direction. This phason excitation is originally the gapless Nambu-Goldstone mode but energy-gapped with the extrinsic pinning frequency ${\omega}_{\rm pin}$ when the helix is subject to commensurate pinning or impurity pinning. Then, depending on the observation frequency ${\omega}_{\rm obs}$ being lower (higher) than ${\omega}_{\rm pin}$, the inductance value $L$ for the phason mode takes a negative (positive) value, whereas the tilting mode contribution always leads to $L$ $ >$ 0~\cite{fref16, fref18}. In reality, the EEMI arising from the spin helix dynamics may be positive or negative, depending on the commensurate/incommensurate modulation, magnetic field, temperature, and ac current amplitude~\cite{fref16, fref17}. In this context, the control of the commensurate- or impurity-pinning effects by the chemical doping procedure in the target materials may be useful to identify the microscopic origin of the emergent inductance.\\
\ \ \ The EEMI of spin helix states was demonstrated experimentally for the first time in the short-period helical magnet ${\rm Gd}_{3}{\rm Ru}_{4}{\rm Al}_{12}$ at low temperatures below 20 K~\cite{fref18}, and found to show large inductance values comparable with commercial coil-inductors in spite of the much smaller $\mu$m-sized device~\cite{fref18}. Recently, the room-temperature emergent induction phenomena have been observed with the spiral magnet ${\rm YMn}_{6}{\rm Sn}_{6}$, whose transition temperature to the spin helical order is beyond room temperature~\cite{fref19}. The magnetic phase diagram [e.g. Fig. 1 (e)] with the progressive magnetic-field-induced variation from screw, transverse conical, fan like and to forced spin-collinear ferromagnetic phases is analogous between the two compounds, ${\rm YMn}_{6}{\rm Sn}_{6}$ and ${\rm Gd}_{3}{\rm Ru}_{4}{\rm Al}_{12}$ with the similar magnetic Kagome lattice. Notably, the magnetic transition temperatures differ by more than one order of magnitude, i.e., 400 K vs. 20 K, and the magnetic modulation direction is different, i.e., along the normal to the Kagome plane in ${\rm YMn}_{6}{\rm Sn}_{6}$ and along the Kagome plane in ${\rm Gd}_{3}{\rm Ru}_{4}{\rm Al}_{12}$, respectively. That the EEMI phenomena are robustly observed in both helimagnetic systems in spite of such dramatic changes in spin helix structure and magnetic energy scale is highly nontrivial, but these facts should be taken as an important implication that the EEMI emerges as a consequence of the general physics of spin helix dynamics. Therefore, in this study we have reinvestigated the detailed phase space occupied by the spin helix using small-angle neutron scattering (SANS), in order to reveal the connection with the EEMI characteristics.
%%SR1%%
In the case of ${\rm Gd}_{3}{\rm Ru}_{4}{\rm Al}_{12}$, it is observed that the negative intrinsic inductance both monotonously and nonlinearly increases in absolute magnitude with the excitation ac current density. In the case of ${\rm YMn}_{6}{\rm Sn}_{6}$, by contrast, a sign change of the inductance to the positive value was observed with changes of temperature, magnetic field and exciting current density; these experimentally observed results imply the variations of the dominant helix dynamic mode, e.g., the phason and the tilting mode, and/or the phason pinning effect. However, control of the sign and magnitude of the emergent inductance, even for a single target material, remains elusive due to such enigmatic temperature- and magnetic-field dependent behaviors of EEMI. The purpose of the present study is to obtain key insights for enhancing EEMI, as well as for controlling its sign around room temperature, by examining the interrelation between various magnetic structures/characteristics and EEMI. For this goal, we attempt to control magnetism systematically by chemical modification of the archetypal high-temperature helimagnet ${\rm YMn}_{6}{\rm Sn}_{6}$. \\
\ \ \ Helimagnetic structures of ${\rm YMn}_{6}{\rm Sn}_{6}$ are formed by the magnetic frustration among exchange interactions along the $c$-axis~\cite{fref20, fref21, fref22, fref23, fref24, fref25, fref26, fref27}. There are two types of nearest-neighbor inter-layer interaction between Mn spins, one is ferromagnetic ${J}_{1} (> 0)$ and the other is antiferromagnetic ${J}_{2} (< 0)$ [Fig.~1(c)]. Only with these ${J}_{1}$ and ${J}_{2}$, the spin texture would be an up-up-down-down like double-antiferromagnetic structure. Here, due to an additional ferromagnetic second-nearest-neighbor coupling ${J}_{3} (> 0)$, a spiral spin texture becomes stable [Fig.~1(d)], whereas the antiferromagnetic-type order is dominant just below the transition temperature~\cite{fref23, fref24, fref25, fref26}. In this study, the moderate chemical modification of the magnetism was done by partially substituting Y with Tb in ${\rm YMn}_{6}{\rm Sn}_{6}$. It is known that the other end compound ${\rm TbMn}_{6}{\rm Sn}_{6}$ hosts a collinear magnetic structure in a whole temperature range below the magnetic transition temperature, and that the magnetic phases change significantly in the solid solution system with ${\rm YMn}_{6}{\rm Sn}_{6}$~\cite{fref28, fref29, fref30, fref31, Mushnikov2015}. We investigate the change in magnetic structure with light Tb substitution (up to 10 $\%$) by means of small angle neutron scattering (SANS) and compare it with the variation of EEMI. 
%%SR2%%
Here, the local effect of Tb substitution on the magnetic interactions is to be noted. The original helical magnetism in ${\rm YMn}_{6}{\rm Sn}_{6}$ ($x = 0$) transforms into the collinear ferromagnetic state upon increasing the Tb content $x$ above 0.4 in $({\rm Y}_{1-x}{\rm Tb}_{x}){\rm Mn}_{6}{\rm Sn}_{6}$, which is due to the strong antiferromagnetic coupling between Mn and Tb~\cite{fref29, fref31}. Therein, upon lowing temperature, the Mn-plane ferromagnetic state with the easy plane anisotropy changes to one with easy-axis ($\parallel c$) anisotropy. Among the inter-plane ($\parallel c$) magnetic exchange interactions, ${J}_{1}$, ${J}_{2}$ and ${J}_{3}$, the ${J}_{2}$ is locally transformed to positive (ferromagnetic) around the doped Tb site as mediated with the strong antiferromagnetic interactions Mn (lower)-Tb and Tb-Mn (upper), which is contrary to the case for ${J}_{2} < 0$ with nonmagnetic Y ion. Furthermore, as described above, the Tb moment shows the easy-axis ($\parallel c$) magnetic anisotropy. Thus, while the global change of the spin helix states remains modest in the case of lightly doped case, e.g., 7 $\%$, the local effect around the doped Tb site occurs through (a) the change of the local ${J}_{2}$ exchange interaction from negative (antiferromagnetic) to positive (ferromagnetic) as well as (b) the local change of the magnetic anisotropy from the easy-plane to the easy-axis type. These features are important in the consideration of the pinning characteristics for the current-driven phason mode relevant to the EEMI~\cite{fref16}.
%%SR2%%

\begin{figure*}
\begin{center}
\includegraphics*[width=6.5in,keepaspectratio=true]{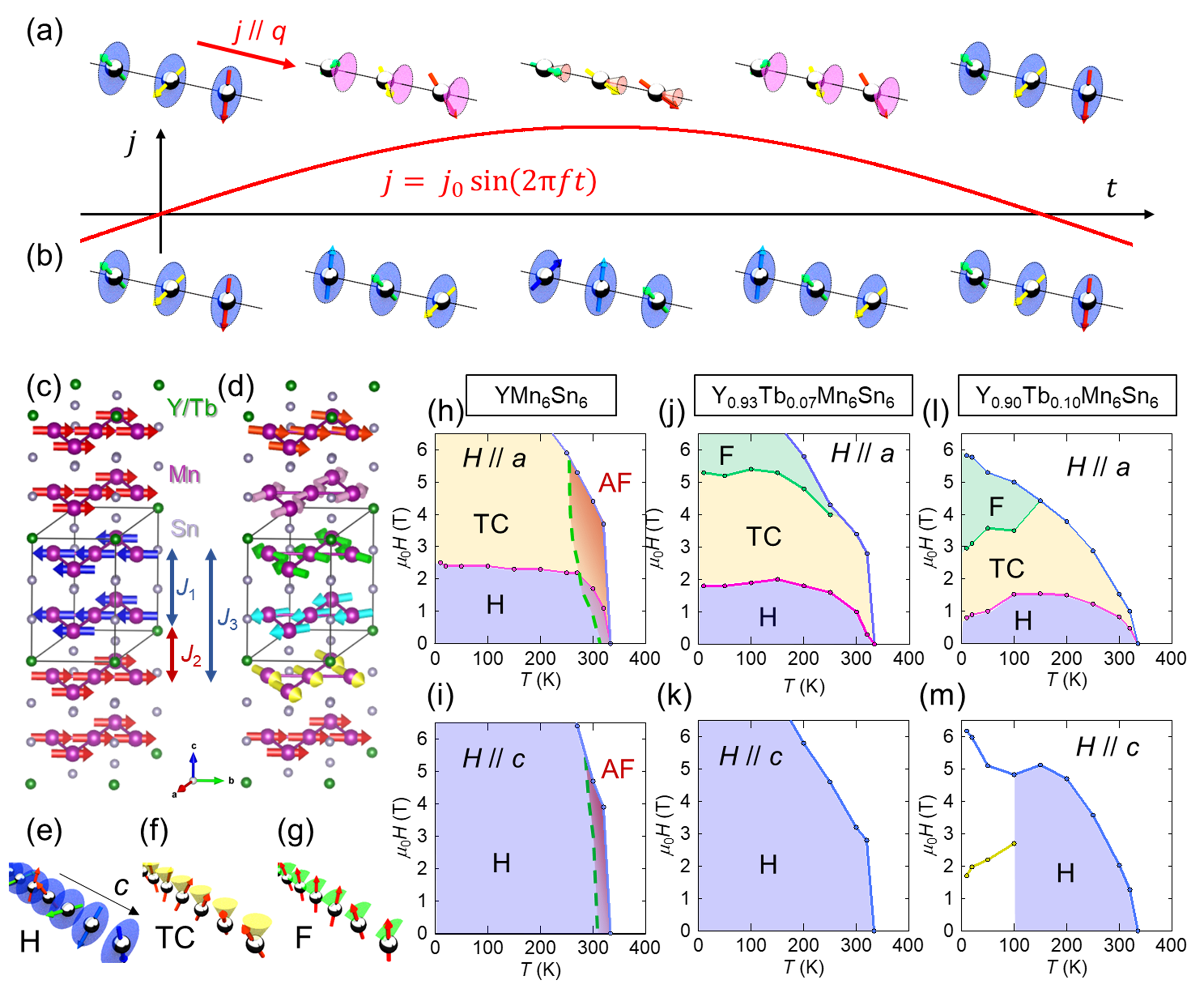}
\caption{
Magnetic phases and phase diagrams in the plane of magnetic field ($H$) and temperature ($T$) of ${\rm (Y, Tb)Mn}_{6}{\rm Sn}_{6}$ in field decreasing process. (a)-(b) Schematic illustrations of (a) spin-tilting mode and (b) phason mode of the spin-helix state responsible for emergent electromagnetic induction during the half cycle of the ac current ($j = {j}_{0}\ {\sin}(2\pi ft)$ excitation. (c)-(d) Interlayer exchange interactions and (c) double antiferromagnetic structure and (d) helical structure on crystal lattice of ${\rm YMn}_{6}{\rm Sn}_{6}$. The illustration was drawn by using VESTA~\cite{fref32}. (e)-(g) Schematic illustrations of (e) proper-screw helical (H), (f) transverse conical (TC) and (g) fan (F) structures. (h)-(m) Overall magnetic phase diagrams of (h), (i) ${\rm YMn}_{6}{\rm Sn}_{6}$, (j), (k)${\rm Y}_{0.93}{\rm Tb}_{0.07}{\rm Mn}_{6}{\rm Sn}_{6}$, and (l), (m)${\rm Y}_{0.90}{\rm Tb}_{0.10}{\rm Mn}_{6}{\rm Sn}_{6}$. The blue, yellow, green, and red regions represent proper-screw helical (H), transverse conical (TC), fan (F), and antiferromagnetic (AF) phases, respectively. The phases below 100 K in ${\rm Y}_{0.90}{\rm Tb}_{0.10}{\rm Mn}_{6}{\rm Sn}_{6}$ under $H \parallel c$ are not identified in the present study.}
\end{center}
\end{figure*}

\section{II. Experiment}
The single crystals of ${\rm (Y, Tb)Mn}_{6}{\rm Sn}_{6}$ were synthesized by Sn-flux method~\cite{fref27}. A mixture of ingredient elements with atomic ratio of (Y, Tb):Mn:Sn = 1:6:30 was put in an evacuated quartz tube and heated to 1050 \(^\circ\)C, subsequently cooled slowly to 600 \(^\circ\)C and then quenched to room temperature. The remaining flux was centrifuged. The single crystallinity was indicated by the well-developed facet structures and was also confirmed by Laue X-ray diffraction. The concentration of Tb was determined by energy dispersive X-ray spectroscopy (EDX). For electric transport measurements, we cut thin plates out of the single crystals by using the focused ion beam (FIB) technique (NB-5000, Hitachi). The thin plates were mounted on silicon substrates with patterned electrodes. We fixed the thin plates to the substrates and electrically connected them to the electrodes by using FIB-assisted tungsten-deposition. We made Au/Ti-bilayer electrode patterns by an electron-beam deposition method.\\
\ \ \ All small-angle neutron scattering (SANS) experiments were carried out using SANS-I instrument at the Swiss Spallation Neutron Source (SINQ), Paul Scherrer Institut, Switzerland, using neutrons with wavelength of either 5 or 6 \AA. The single-crystalline samples were attached onto an Al plate holder and loaded into a cryomagnet installed at the sample position of the beamline. For measurements with $H \parallel a$ ($H \parallel c$), the magnetic field was applied nearly parallel (perpendicular) to the incident collimated neutron beam. The diffracted neutrons were collected by a two-dimensional multidetector placed 1.85 m behind the sample, and translated 0.46 m horizontally in the plane perpendicular to the incoming beam direction in order to access an extended $q$-range along $c$*. The diffraction measurements were done by recording SANS patterns over a range of sample-cryomagnet ensemble rotation angles that were sufficient to move the magnetic peaks through the Bragg condition at the detector.\\
\ \ \ Magnetic-field dependence of complex resistivity was measured with use of lock-in amplifiers (SR-830, Stanford Research Systems). We input a sine-wave current and recorded both in-phase (Re $V$) and out-of-phase (Im $V$) voltage with a standard four-terminal configuration. Background signals were estimated by measuring a short circuit where the terminal pads were connected by Au/Ti-bilayer electrode patterns. We subtracted the background signals from the measured data. Magnetization was measured by using Quantum Design PPMS-14 T with ACMS option.

\section{III. Overall magnetic phase diagrams of ${\bf (Y, Tb)}{\bf Mn}_{\bf 6}{\bf Sn}_{\bf 6}$}
The magnetic phase diagrams of ${\rm (Y, Tb)Mn}_{6}{\rm Sn}_{6}$ were determined by magnetization and small angle neutron scattering (SANS) measurements. In the known case of ${\rm YMn}_{6}{\rm Sn}_{6}$~\cite{fref22, fref23, fref24, fref25}, an incommensurate helical state (H [Fig.~1(e)]) with the wavevector (${\bm q}$) parallel to the $c$-axis is stable at zero magnetic field below 330 K. As increasing the magnetic field applied parallel to the $a$-axis ($H \parallel a$), the spin texture changes to a transverse conical state (TC [Fig.~1(f)]) and then to a fan state (F [Fig.~1(g)]) at above 7 T below 150 K. With further increasing $H \parallel a$, the state reaches a forced ferromagnetic state (FF). With the magnetic field parallel to the $c$-axis ($H \parallel c$), on the other hand, the helical state is continuously deformed to a longitudinal conical state and then a forced ferromagnetic state at higher fields. The spiral states always propagate along the $c$-axis. Above 250 K, the magnetic periodicity becomes commensurate with ${\bm q}_{\rm C}$ = (0 0 0.5), namely the up-up-down-down like double antiferromagnetic state (AF) is stabilized near the magnetic phase boundary. From the previous study on polycrystalline samples~\cite{fref28, fref29, fref30, fref31, Mushnikov2015}, when the Tb concentration is less than 10 $\%$, it is expected that there will be no drastic change in the magnetic phase at zero magnetic field. This time, we targeted the compounds with 0 $\%$, 7 $\%$, and 10 $\%$ concentrations of Tb dopant on Y-site in ${\rm (Y, Tb)Mn}_{6}{\rm Sn}_{6}$. The samples are single crystals with fairly homogeneous (Y,Tb) compositions grown by the Sn flux method. 7 $\%$ is the low Tb concentration limit for high quality samples that could be synthesized by the flux method. For smaller Tb concentrations, a phase separation was observed to occur.
Figures~1(h)-(m) show the magnetic phase diagrams of the respective compositions as deduced in the magnetic-field descending procedure. The phase boundaries were determined on the basis of magnetization curves (see also Figs.~S1-S3 of Supplemental Material~\cite{SM}). As for the AF phase, the phase boundary cannot be clearly defined from the $M$-$H$ curves, and instead it is determined by the SANS measurements as described later. Previous neutron diffraction studies have revealed that the magnetic moment of Tb is antiferromagnetically coupled to that of Mn~\cite{fref31}. This assignment is consistent with the present magnetization measurements; the saturation magnetization at low temperature (10 K) is in accord with the value based on the assumption that Mn spins (2.1 ${\mu}_{\rm B}$/atom) and Tb moments (9.0 ${\mu}_{\rm B}$/atom) are antiferromagnetically coupled. This antiferromagnetic coupling is robustly sustained up to at least 14 T. When the Tb concentration is below 10 $\%$, the magnetic ordering temperature does not change significantly in the range of 330-335 K. For Tb 10 $\%$ doping, an additional faint peak on the $M$-$T$ curves was observed around 100 K (see also Fig.~S3 of Supplemental Material~\cite{SM}). The magnitude of the transition magnetic field decreases as the amount of Tb increases, regardless of the direction of the magnetic field. Since the behavior of the transition under a magnetic field is similar at each concentration, it can be reasonably inferred that the magnetic structure of the Mn spin shows similar temperature-magnetic field dependence for these undoped and Tb-doped crystals, as shown in Figs.~1(h)-(m).

\section{IV. Magnetism of ${\bf (Y, Tb)}{\bf Mn}_{\bf 6}{\bf Sn}_{\bf 6}$ revealed by small angle neutron scattering}
\subsection{A. ${\bf Y}{\bf Mn}_{\bf 6}{\bf Sn}_{\bf 6}$}
Figure 2(a) shows the setup in the SANS measurement. A neutron beam propagates nearly parallel to the $a$-axis of the single crystal sample and hence nearly perpendicular to the reciprocal $a$*-$c$* plane. The horizontal (vertical) direction of the two-dimensional detector corresponds to the $c$* ($a$*) direction. No diffraction spots were observed along the $a$*-direction at any temperature and magnetic field condition. Figure 2(b) shows SANS patterns obtained at zero magnetic field and various temperatures. Diffraction peaks are observed at three different wavevectors; (i) the incommensurate wavevector ${\bm q}_{\rm IC}$ corresponding to helical spiral states, (ii) the commensurate ${\bm q}_{\rm C}$ = (0 0 0.5) corresponding to the AF structure, and (iii) (0 0 1)$ -{\bm q}_{\rm IC}$. While the two kinds of inter-layer distances between neighboring Mn planes of ${\rm YMn}_{6}{\rm Sn}_{6}$ are similar to each other, the interplane rotation angle of the ferromagnetically-aligned in-plane moment must be different between the pair coupled via ferromagnetic ${J}_{1}$ and the pair coupled via antiferromagnetic ${J}_{2}$. The spirals of the second neighbor Mn spins along $c$-axis have a nearly constant angle of rotation. This double spiral state produces (0 0 1)$ -{\bm q}_{\rm IC}$ spots, and the nonmonotonicity of the rotation angle is reflected in the intensity ratio of (0 0 1)$ -{\bm q}_{\rm IC}$ and ${\bm q}_{\rm IC}$ (see also Fig.~S4 of Supplemental Material~\cite{SM}). At 10 K the helical period is about 3.3 nm. As the temperature rises, it shortens to 2.2 nm at 320 K, a temperature where the helical and AF states are found to coexist. Figure~2(c) shows the diffraction profile at each temperature. The H state is dominant at low temperatures, and the intensities of the peaks at ${\bm q}_{\rm IC}$ and ${\bm q}_{\rm C}$ become comparable near the transition temperature (320 K). In detail, the peak of ${\bm q}_{\rm IC}$ is known to be composed of multiple spirals with close periods ~\cite{fref24, fref26}. Multiple peaks are observed in the present SANS profiles as well; here the position and intensity are fitted with a single Gaussian peak [Fig.~2(d)] and discussed as a whole hereafter. We note that such a discommensuration-like feature may contribute to the dynamics of the spin texture and enhance the emergent inductance discussed later. \\
\begin{figure*}
\begin{center}
\includegraphics*[width=6in,keepaspectratio=true]{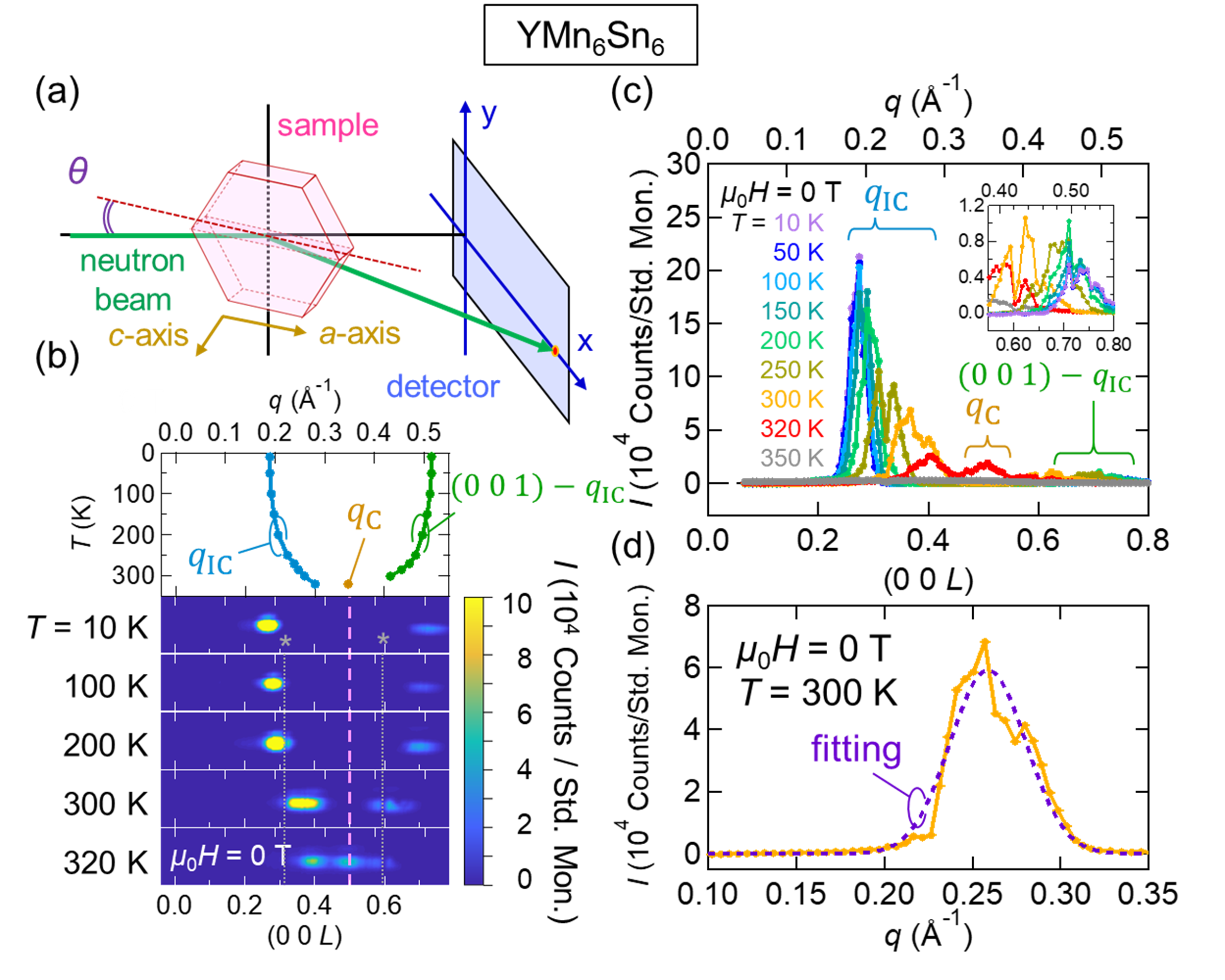}
\caption{SANS profiles of ${\rm YMn}_{6}{\rm Sn}_{6}$ at zero magnetic field. (a) Setup and configuration of small angle neutron scattering (SANS). (b) SANS patterns and positions of diffraction spots along the $c$*-axis at 0 T. The upper horizontal scale is for the propagation vector $|{\bm q}|$ value in ${\rm \AA}^{-1}$ and the lower scale for the reciprocal lattice unit. The pink dashed line corresponds to the commensurate $L$ = 0.5 position. Vertical dotted lines with asterisk indicate blind spot positions on the present SANS detector; when the diffraction spot coincides with the blind spot, the diffraction intensity is dumped as an artifact. (c) SANS profiles along $c$*. The inset shows a magnified view of higher angles. (d) The SANS profile of incommensurate ${\bm q}_{\rm IC}$ at 300 K. The observed profile (orange) is not a sharp single peak, but more likely composed of multiple peaks. The purple dashed line indicates the result of fitting to a single Gaussian for a course-grain analysis.}
\end{center}
\end{figure*}
\ \ By tracking the temperature and magnetic field dependence of ${\bm q}_{\rm IC}$ and ${\bm q}_{\rm C}$, the spiral and AF phases are clarified in the temperature-magnetic field plane. The development of the SANS intensities for $H \parallel a$ is summarized as an intensity contour map in Fig.~3(a) for the helical spiral order described by ${\bm q}_{\rm IC}$, and in Fig.~3(b) for the AF structure described by ${\bm q}_{\rm C}$.
It is known that the magnetic field along the $a$-axis stabilizes the AF state near the transition temperature (330-335 K) ~\cite{fref24, fref26}. The present study confirms that the AF state remains in a magnetic field along the $c$-axis as well. Figures 3(c) and 3(d) show the distribution of each SANS spot intensity for $H \parallel c$, while the clear coexistence region of H+AF is rather narrow. The main magnetic phase changes from spiral to AF around the high-temperature phase boundary. Here all the presented magnetic field dependent data are obtained in a magnetic-field descending process at fixed temperatures. Figures 3(e) and 3(f) show SANS profiles at 300 K for several magnetic fields. In general, the signal strength of spin modulation decreases as the magnetic field increases because of the tilt of spins toward the external magnetic field. From the above results, we confirm that the magnetic phase of ${\rm YMn}_{6}{\rm Sn}_{6}$ falls into the valley of AF phase when ramping the field down from the FF phase at high temperature. This phenomenon is independent of the magnetic field direction, $H \parallel c$ or $H \parallel a$. 
\begin{figure*}
\begin{center}
\includegraphics*[width=6.5in,keepaspectratio=true]{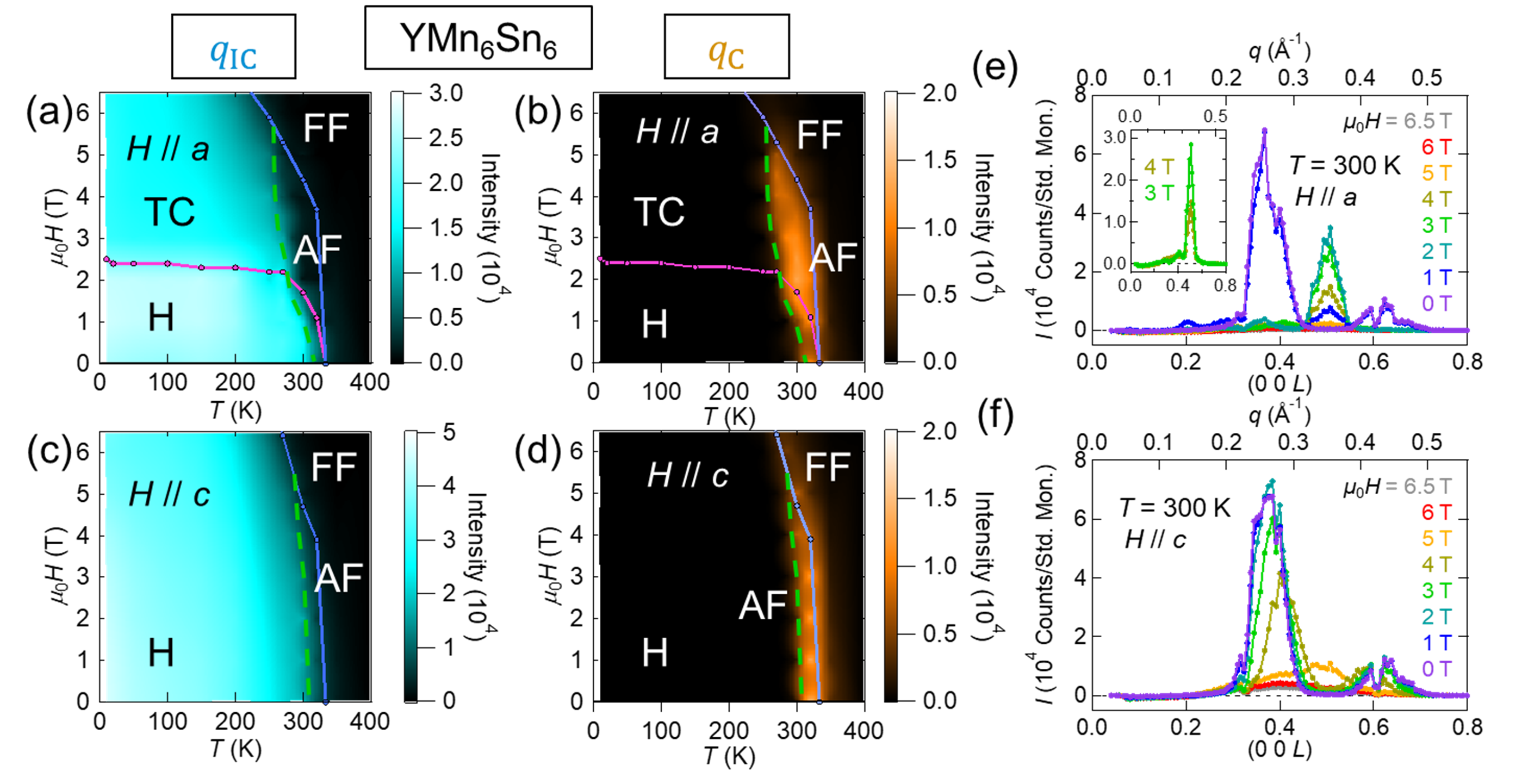}
\caption{Magnetic field dependence of SANS magnetic satellite in ${\rm YMn}_{6}{\rm Sn}_{6}$. (a)-(d) Color maps summarizing intensities profiles of (a), (c) incommensurate (${\bm q}_{\rm IC}$) spiral and (b), (d) antiferromagnetic (${\bm q}_{\rm C}$) states in magnetic fields along (a), (b) the $a$-axis and (c), (d) the $c$-axis. (e)-(f) SANS profile along the $c$*-axis at 300 K at various magnetic fields along (e) the $a$-axis and (f) the $c$-axis. The inset of (e) shows a magnified view of profile at magnetic fields of 3 T and 4 T.}
\end{center}
\end{figure*}

\subsection{B. ${\bf Y}_{\bf 0.93}{\bf Tb}_{\bf 0.07}{\bf Mn}_{\bf 6}{\bf Sn}_{\bf 6}$ ($\bf Tb7$ $\%$) and ${\rm Y}_{\bf 0.90}{\bf Tb}_{\bf 0.10}{\bf Mn}_{\bf 6}{\bf Sn}_{\bf 6}$ ($\bf Tb 10$ $\%$)}
Using the same experimental setup as for ${\rm YMn}_{6}{\rm Sn}_{6}$, the magnetic structures of the Tb-doped compounds, ${\rm Y}_{0.93}{\rm Tb}_{0.07}{\rm Mn}_{6}{\rm Sn}_{6}$ (Tb 7 $\%$) and ${\rm Y}_{0.90}{\rm Tb}_{0.10}{\rm Mn}_{6}{\rm Sn}_{6}$ (Tb 10 $\%$), were also examined by SANS measurements. Figures 4(a) and 4(b) show SANS patterns at zero magnetic field and at various temperatures for the Tb 7 $\%$ and Tb 10 $\%$ compounds, respectively. It is clear that the AF peak locating at ${\bm q}$ = (0 0 0.5) is totally extinguished by the Tb doping. The temperature dependence of the observed wavevectors of the diffraction spots is summarized in Fig.~4(c) together with the corresponding result for the undoped (Tb 0 $\%$) crystal. The magnetic period ${\it \lambda}$ (= 2$\pi$/$q$) in the H phase of the Tb 7 $\%$ doped crystal is comparable or slightly longer in comparison with the H phase (2.2 nm $\leq$ ${\it \lambda}$ $\leq$ 3.3 nm) of the undoped (Tb 0 $\%$) one. For the Tb 10 $\%$ sample the spiral period is clearly longer than the undoped compound, varying from 3.7 nm $\leq$ ${\it \lambda}$ $\leq$ 4.1 nm and with a weaker temperature-dependence. It appears from the change of ${\it \lambda}$ that the contribution from the ferromagnetic interaction is strengthened via the modification of ${J}_{2}$ by Tb substitution. Even near the transition temperature, the position of ${\bm q}_{\rm IC}$ is far from the commensurate (0 0 0.5), and there is no diffraction spot corresponding to the AF (${\bm q}_{\rm C}$) state in any temperature-field region. Incidentally, the peak at (0 0 1)$ -{\bm q}_{\rm IC}$, is barely observed for Tb 7 $\%$ around (0 0 0.8), which corresponds to the upper limit of detectable $q$ range in the present SANS setup, and beyond the detectable $q$-range for Tb 10 $\%$. \\
\ \ \ Another feature in the SANS data from the Tb 10 $\%$ sample that differs from ${\rm YMn}_{6}{\rm Sn}_{6}$ is the presence of weaker broad peaks on the lower $q$ side of the main incommensurate peak as can be seen in Fig.~4(b). This broad peak shifts to a further lower $q$ as the temperature increases. Such additional H states, composed of multiple peaks with close $q$ values, may be induced by the further magnetic frustration added by the Tb substitution. \\
\begin{figure*}
\begin{center}
\includegraphics*[width=7in,keepaspectratio=true]{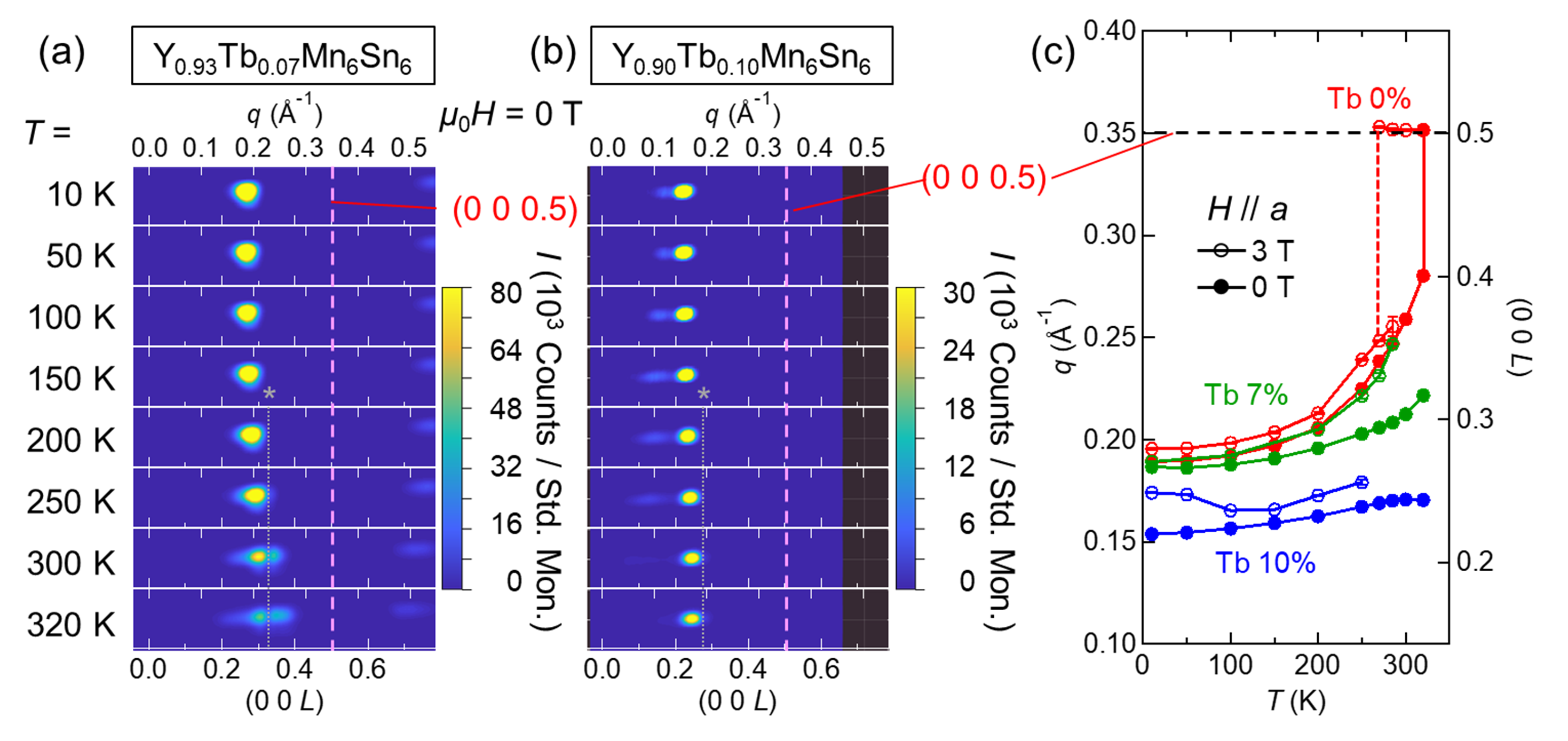}
\caption{SANS profiles along (0 0 $L$) of Tb doped ${\rm YMn}_{6}{\rm Sn}_{6}$ crystals at zero magnetic field, (a) for ${\rm Y}_{0.93}{\rm Tb}_{0.07}{\rm Mn}_{6}{\rm Sn}_{6}$ (Tb 7 $\%$) and (b) for ${\rm Y}_{0.90}{\rm Tb}_{0.10}{\rm Mn}_{6}{\rm Sn}_{6}$ (Tb 10 $\%$). The pink dashed lines correspond to the commensurate (0 0 0.5), at which no peak is discerned. Vertical dotted lines with asterisk indicate the blind spot position of the present SANS detector; when the diffraction spot coincides or overlaps with the blind spot, the diffraction intensity is dumped as an artifact. (c) Temperature dependence of the spiral wavevector $q$ values for Tb undoped, 7 $\%$, and 10 $\%$ doped crystals at zero field and at a magnetic field of 3 T applied along the $a$-axis. The Tb undoped crystal shows the transition from incommensurate (${\bm q}_{\rm IC}$) helical (H, at 0 T) or transverse-conical (TC, at 3 T) to commensurate (${\bm q}_{\rm C}$) antiferromagnetic (AF) state, as shown by vertical dashed (0 T) and solid (3 T) lines.}
\end{center}
\end{figure*}
\ \ By using SANS in various magnetic fields, we have revealed the spiral magnetic phases for the lightly Tb-doped crystals. Figs.~5(a) and 5(b) exemplify the SANS profiles of the ${\bm q}_{\rm IC}$ diffraction for the Tb 7 $\%$ and Tb 10 $\%$ crystals, respectively, at 300 K under various magnetic fields applied along the $a$-axis. There is only slight change in the magnetic period ($q$ value) with applied magnetic field; see also Fig.~4(c). In Figs.~5(c) and 5(d), the ${\bm q}_{\rm IC}$ diffraction intensity is plotted on the magnetic phase diagrams based on the magnetization measurements [Figs.~1(h)-(m)]. The incommensurate magnetic modulations are observed in the whole region below the ferromagnetic transition magnetic field. From these results, we have revealed that low-concentration Tb substitution tends to slightly expand the magnetic period of the spiral structures while totally eliminating the AF phase around the phase boundary. The thermal fluctuation increased by elevating temperature is known to cause the incommensurate to commensurate (e.g., AF) crossover of the magnetic structure, due perhaps to the spin-lattice interaction or magnetostriction effect, as reported for other magnetically frustrated systems~\cite{fref33}. In that case, the introduction of the pinning centers like the present Tb doping appears to lead to proliferation of discommensurations and hence to destroy the commensurate order while nonetheless sustaining the originally stable incommensurate state.
 
\begin{figure*}
\begin{center}
\includegraphics*[width=5in,keepaspectratio=true]{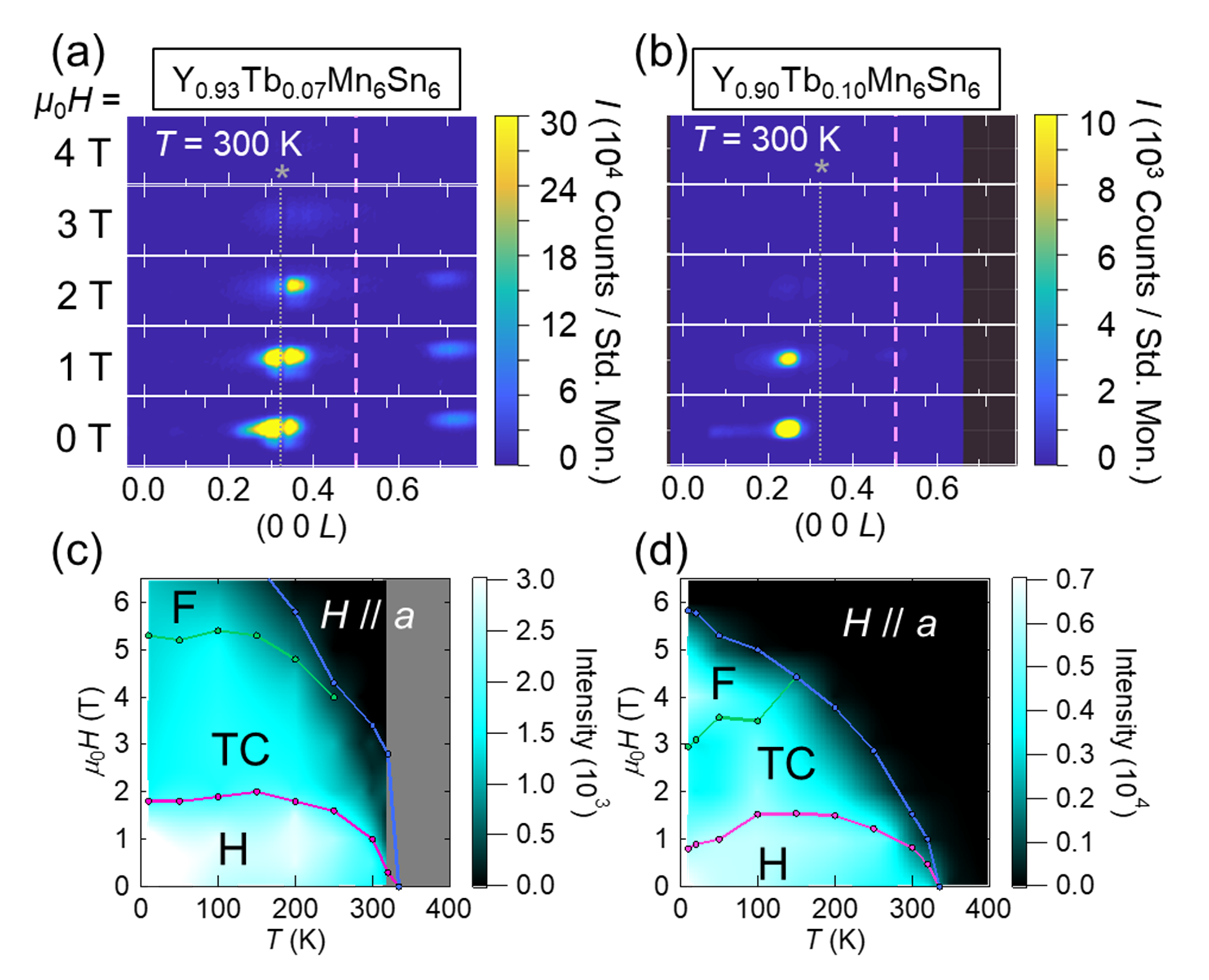}
\caption{SANS results of ${\rm Y}_{0.93}{\rm Tb}_{0.07}{\rm Mn}_{6}{\rm Sn}_{6}$ (Tb 7 $\%$) ${\rm Y}_{0.90}{\rm Tb}_{0.10}{\rm Mn}_{6}{\rm Sn}_{6}$ (Tb 10 $\%$) under various magnetic fields applied along the $a$-axis. (a)-(b) SANS profiles of the incommensurate spiral state (${\bm q}_{\rm IC}$) under magnetic fields at 300 K, for (a) Tb 7 $\%$ and (b) Tb 10 $\%$ doped crystals. Vertical dotted lines with asterisk indicate the blind spot position of the present SANS detector; when the diffraction spot coincides with the blind spot, the diffraction intensity is dumped as an artifact. (c)-(d) The SANS intensity contour maps on the magnetic phase diagrams for (a) Tb 7 $\%$ and (b) Tb 10 $\%$ doped crystals.}
\end{center}
\end{figure*}

\section{V. Emergent electromagnetic inductance (EEMI) of ${\bf (Y, Tb)}{\bf Mn}_{\bf 6}{\bf Sn}_{\bf 6}$}

Keeping in mind the above-described Tb substitution-induced modification of the magnetic structure, we turn to discuss the effect of Tb-substitution on emergent electromagnetic inductance (EEMI) of ${\rm YMn}_{6}{\rm Sn}_{6}$. The emergent inductance value $L$ is directly related to the imaginary part of the ac electric resistivity Im ${\rho}$ via the relation,\\
\begin{eqnarray*}
L = {\rm Im}\rho\ d/2\pi fS
\end{eqnarray*}
\ \ \ Here, $d$ is the distance between electrodes on the sample, $S$ is the cross-sectional area of sample, and $f$ is the ac current frequency. We fabricated micro-scale devices with the dimension of $d$ = 25-35 $\mu$m and $S$ = (2-3) $\mu$m $\times$ (8-10) $\mu$m by the focus-ion-beam (FIB) method. Figures~6(a)-(l) show the magnetic field ($H \parallel a$ and $H \parallel c$) dependence of Im ${\rho}$ at various temperatures (200 K and 300 K) and Tb concentrations (0 $\%$, 7 $\%$, and 10 $\%$), measured using the ac input current density $j = {j}_{0}\ {\sin}(2\pi ft)$ (${j}_{0} =\ $2.5$\times {10}^{4}$ A/${\rm cm}^{2}$, $f$ = 500 Hz, $j \parallel c$). All devices used in the experiment showed larger signals than the false background signals, which come from parasitic impedance and signal delay between current source and detector. In the case of ${\rm YMn}_{6}{\rm Sn}_{6}$ [Tb 0 $\%$, Figs.~6(a)-(d)], a negative inductance appears in the H phase at zero magnetic field, in accord with the previous result~\cite{fref19}. With this current density, the origin of the negative inductance has been assigned mainly to the current-induced phason motion [Fig.1b] with the extrinsic pinning frequency (a few kHz above the observation frequency~\cite{fref19}. The large enhancement of the negative inductance around the H and TC coexisting region for $H \parallel a$ [Fig.~6(a)] is due perhaps to the H-TC domain-wall (DW) motions driven by ac current, as a sort of phason-like motion, while the extrinsic pinning frequency should be still above the observation frequency for this DW state. At a higher temperature, e.g., 300 K, the inductance turns to positive with the increase of magnetic field. In a magnetic field along the $a$-axis, a sign change occurs upon entering the TC state from the H state, while in a magnetic field along the $c$-axis, a positive peak structure appears near the ferromagnetic transition where the AF state coexists with the H (to be precise, longitudinal conical) state.\\

\subsection{A. Doping-induced pinning effect on negative EEMI}
\ As for the Tb 7 $\%$ doping, the negative inductance is significantly suppressed in comparison with the pristine (Tb 0 $\%$) case, while the positive inductance survives quantitatively [Figs.~6(e)-(h)]. Such a large impact of the Tb doping confirms that the positive and negative inductance components have different microscopic origins, as respectively assigned to the tilting [Fig.~1(a)] and phason [Fig.~1(b)] motions of the spin spiral~\cite{fref16}. Namely, the extrinsic pinning effect on the phason motion strengthened by Tb doping likely leads to the critical suppression of the negative component of the EEMI that is dominant in the H phase of the pristine (Tb 0 $\%$) compound. By contrast, the tilting motion is less affected by the pinning effect, and hence the positive component of the emergent inductance is largely preserved.\\
%%SR5%%
\ \ \ To gain insight into the pinning mechanism of the phason mode, an important point that needs to be taken into account is the difference between the spin-spiral state (the present case) and the conventional collinear spin density wave (SDW). In the latter case, the phason is readily pinned by the impurity, showing the finite extrinsic pinning frequency of its spectrum. In sharp contrast, the phason in the incommensurate spin spiral state with the constant spin moment amplitude remains gapless even under the disorder, as long as the spin rotational symmetry of the Hamiltonian is kept intact. To gap the phason mode, needed is a coupling between the impurity (Tb) perturbation and the more or less elliptic, not perfectly circular, helical modulation of the spin moment amplitude, which means the admixture of the sinusoidal (SDW-like) moment modulation into the helical spin moment. In contrast to the case of rare-earth 4$f$ moments, some ellipticity of the helix or resultant local charge density modulation is quite plausible for 3$d$ Mn moments that also contribute to the conduction band formation as in the conventional SDW metals (e.g., chromium~\cite{Fawcett1994}). On one hand, even in the presence of local modulations of the exchange interaction (a) and the magnetic anisotropy (b), as long as the U(1) symmetry in spin space remains like the present Tb-doped case, the U(1) rotation of the whole spins is still gapless. This corresponds to the usual magnon together with the tilting mode (${m}_{z}$: the uniform magnetization perpendicular to the spin rotation place) as the canonical conjugate generator. On the other hand, in the presence of the elliptic helical modulation, the phason, which is separated from the spin rotation mode, is pinned due to impurities. These characteristic features of the phason mode in the helix may explain why the EEMI measurement on the nominally pure crystal of ${\rm YMn}_{6}{\rm Sn}_{6}$ indicates the ultra-low extrinsic pinning frequency $f$ $\leq$ 10 kHz~\cite{fref18, fref19}. In contrast, the intentional doping of Tb likely escalates the extrinsic pinning frequency to become much higher than the observation frequency ($f$ $\leq$ 1 kHz) and suppresses the negative EEMI signal. In a future study, it is desirable to experimentally verify the correlation between this ellipticity of the spin helix and the frequency characteristics of EEMI.\\
\ \ \ The rather moderate effect of Tb doping observed for the positive EEMI at relatively high temperatures can also be explained by the fact that the tilting mode ${m}_{z}$, which is responsible for the positive EEMI, is already subject to the influence of the easy plane anisotropy energy ${Km}_{z}^{2}$ and hence rather insensitive to the presence of the magnetic (Tb) impurities.
%%SR5%%
When the Tb concentration is further increased to 10 $\%$, the negative inductance is almost completely eliminated, while the reduction of the positive inductance is less significant, remaining at the level of about 20 $\%$ [Fig.~6(i)-(l)]. A possible mechanism for the reduced positive emergent inductance (tilting mode) may stem from the static local canting of Mn moment off the spiral plane induced by Tb doping, which as magnetic disorder may cause the suppression of the current-induced tilting mode dynamics. One other mechanism to reduce the EEMI with Tb doping is the elongation of magnetic period $\lambda$ as seen upon Tb 10 $\%$ doping; e.g., from 2.4 nm (IC) or 1.8 nm (AF) for Tb 0 $\%$ to 3.7 nm (IC) for Tb 10 $\%$ at 300 K, see Fig. 3(c). However, the reduction of magnitude of negative emergent inductance is by far larger than that expected from the change in ${\it \lambda}$ (= 2$\pi$/$q$); Im $\rho \propto q$. Thus, the other important mechanism, namely the pinning-induced suppression of the phason mode, plays a dominant role for the drastic reduction of negative EEMI. By contrast, the positive inductance component is relatively preserved and its qualitative magnetic field dependence is commonly observed irrespective of the Tb concentration ($\leq$ 10 $\%$). To be noted here is that there is no AF phase even at 300 K in the cases of Tb 7 $\%$ and 10 $\%$, as opposed to the case of ${\rm YMn}_{6}{\rm Sn}_{6}$ (see Fig.~5). Therefore, the origin of the positive inductance common to the pristine and Tb-doped compounds is not an exclusive characteristic of the AF phase itself. In other words, the current-induced tilting motion of the AF state at relatively high temperatures (e.g., 300 K) may be responsible for the positive emergent inductance as in the other incommensurate TC state near the phase boundary (see Fig.7(b)).\\
%%SR4%%

\subsection{B. Domain wall EEMI}
In common with the Tb 7 $\%$ and 10 $\%$ crystals, the sharp positive peak anomalies of the inductance are observed at the phase boundary between H and TC for $H \parallel a$ or between H and FF for $H \parallel c$, as typically seen in Figs. 6(e), (g), (h), (i), (k) and (l). These may indicate the current-driven free or depinned motion of DWs in analogy to the gapless or depinned phason mode, which can give the positive sign of EEMI~\cite{fref16, fref19}. On the contrary, as seen in Fig. 6(a), in the field region between the H and AF phases for Tb 0 $\%$, there is observed a rather broad negative EEMI peak. In this case, the microscopic coexistence of the TC and AF domains viewed as accumulated DWs is likely to be the origin of the enhanced negative EEMI response. The negative sign of EEMI observed therein indicates the bound motion of DW due to pinning at a relatively low ac current density, ${j}_{0} =\ $2.5$\times {10}^{4}$ A/${\rm cm}^{2}$. Later (in V-D), we note the current-induced depinning transition of such a DW phason-like mode accompanied by the change of EEMI toward the positive value.\\
\ \ \ Incidentally, a discrepancy between the inductance peak field and the phase boundary field is sometimes discerned, for example, in Fig. 6(g). This is due perhaps to the slight (within $\pm$0.7 $\%$) deviation of the Tb stoichiometry of the micro-device samples from that of the corresponding bulk crystals, on which the magnetization measurements were done for deducing the phase diagrams [Figs. 1(j), 1(k), 1(l) and 6(m)].  As seen in Figs. 1(k) and (m), for example, the H-FF phase-transition fields steeply decrease with Tb doping, crudely at a rate of ~0.4 T per Tb 1$\%$ at 300 K. The discrepancy between the magnetic field values for the sharp EEMI signal and the phase boundary is of this order, and we speculate that the DW EEMI signal field-position sensitively reflects the variation of the Tb content in the Tb 7 $\%$ (averaged value) crystal.
%%SR4%%
%%SR3%% 
\begin{figure*}
\begin{center}
\includegraphics*[width=7in,keepaspectratio=true]{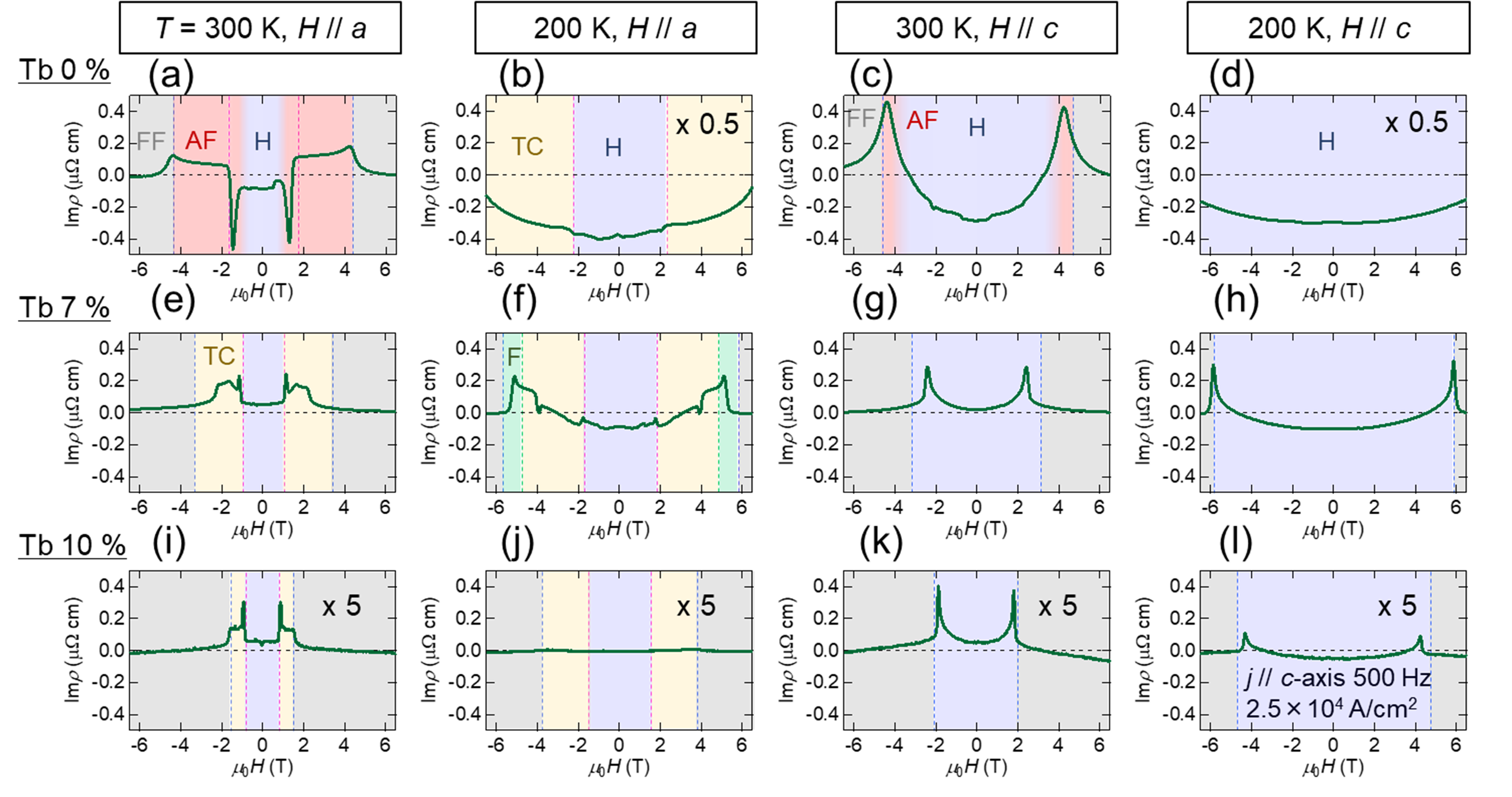}
\caption{(a)-(l) Magnetic field ($H \parallel a$, $H \parallel c$) dependence of emergent electromagnetic induction with variations of Tb-concentration (0 $\%$, 7 $\%$, and 10 $\%$) and temperatures (200 K and 300 K) for ${\rm (Y, Tb)Mn}_{6}{\rm Sn}_{6}$. The imaginary part of the complex resistivity Im $\rho$ as the materials quantity representing the inductance was measured with an ac input current density $j = {j}_{0}\ {\sin}(2\pi ft)$ (${j}_{0} =\ $2.5$\times {10}^{4}$ A/${\rm cm}^{2}$, $f$ = 500 Hz, $j \parallel c$). The blue, yellow, red, green and gray shadows represent proper-screw helical (H), transverse conical (TC), antiferromagnetic (AF), fan (F) and forced ferromagnetic (FF) phases, respectively.}
\end{center}
\end{figure*}

\subsection{C. Doping effect on positive EEMI}
Figures 7(a)-(f) show the color contour maps of Im $\rho$ overlaid on the magnetic phase diagrams. It is clear that with the increase of Tb doping the negative component (shown with blue color shading in the figures) in the H and TC states are rapidly and conspicuously suppressed. By contrast, the positive inductance (shown in red shading) appears around the high-temperature phase boundaries and persists robustly against Tb doping, and irrespective of the dominant magnetic modulation, i.e., commensurate (AF) or incommensurate (H) state. This observation points to the importance of the local spin dynamics, rather than the static form of magnetic order, as the origin of the positive inductance. Spin fluctuations are anticipated to be large around high-temperature phase boundary. These enhanced magnetic fluctuations and their response to the ac current are likely to play a part in the origin of the positive inductance. When the AF state is viewed in spin projective space, only the two points at the north pole and the south pole are occupied. Even when it is driven by current, it dynamically sweeps only a line with zero-solid angle, and hence does not contribute to the generation of emergent electric field. In the presence of thermally enhanced spin fluctuations, however, the AF state can cover a finite area around the poles in the spin projective space, which can result in emergent electric fields when driven by current. This scenario is in accord with the mechanism of the positive emergent inductance based on tilting motion, while the phason-like motion as the origin of negative inductance is suppressed in such a thermally disordered state. In this context, the possible gap of the phason in the commensurate AF state present in the pristine (Tb 0 $\%$) crystal may play some role in suppressing the phason excitations (which give the negative component of emergent inductance) to make the tilt excitations (which give the positive component) more dominant. Nevertheless, the thermal fluctuations, irrespective of commensurate or incommensurate spin orders, are likely to be more important, judging from the common behavior of the positive inductance signals irrespective of the Tb-doping levels 0 $\%$ or 7 $\%$. It is also to be noted here that the present commensurate (AF) phase does not show a simple up-down-up-down spin configuration, but instead the up-up-down-down type along the $c$-axis; intuitively, the latter configuration appears more favorable to host the dynamically noncoplanar configuration with assist of thermal fluctuation. This scenario of the thermal-fluctuation-induced generation of emergent induction needs to be verified in spin-collinear magnets without any adjacent spiral phases, and analyzed in a more quantitative way by elaborate model simulations.\\
\begin{figure*}
\begin{center}
\includegraphics*[width=7in,keepaspectratio=true]{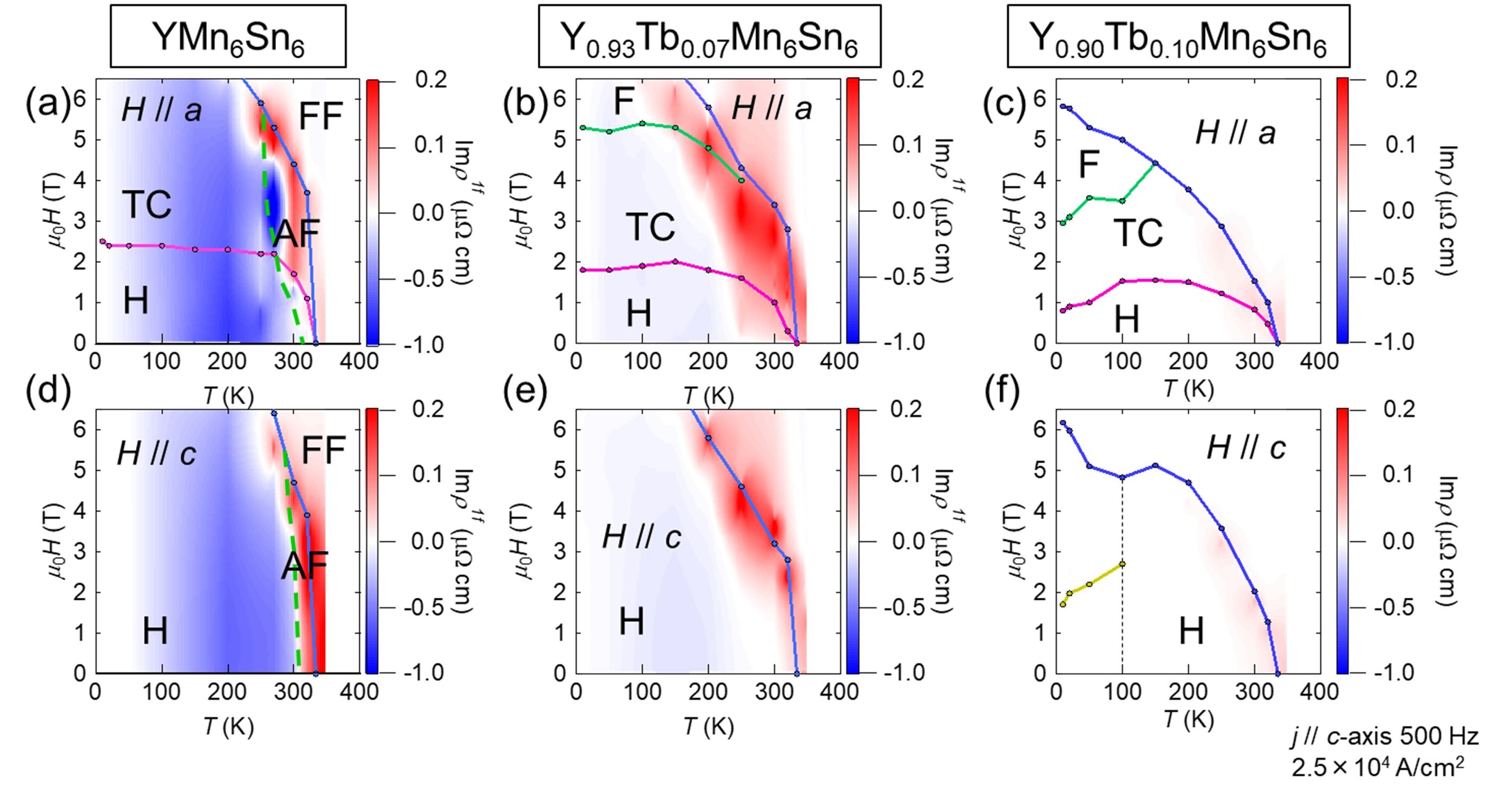}
\caption{(a)-(f) Color maps of Im $\rho$ at ${j}_{0} =\ $2.5$\times {10}^{4}$ A/${\rm cm}^{2}$, $f$ = 500 Hz in the $T$-$H$ plane at each Tb concentration and magnetic field orientation. Red and blue areas correspond to positive and negative Im $\rho$, respectively. The color depth corresponds to the magnitude of the value; see the color scale bar.}
\end{center}
\end{figure*}

\subsection{D. Current-nonlinear EEMI}
Lastly, we investigate the nonlinear current-density dependence of the observed emergent inductance. Figs.~8(a)-(f) show the magnetic field dependence of Im $\rho$ when the current density is varied up to ${j}_{0} =\ $5.0$\times {10}^{4}$ A/${\rm cm}^{2}$ for each Tb composition with $H \parallel a$. Figure~8(g) is a plot of Im $\rho$ as a function of current density at some selected fixed magnetic fields. The magnitude of Im $\rho$ generally increases with the increase of current density, showing a strongly nonlinear behavior, in particular, in the region of the negative inductance value, e.g., for the Tb 0 $\%$ compound at 200 K and 0 T, where the phason mode contribution is dominant. For most conditions, a higher current density leads to a monotonous increase of the absolute magnitude of Im $\rho$. For the Tb 0 $\%$ compound at 300 K and 1.5 T, however, Im $\rho$ shows a nonmonotonic behavior with respect to current density; first becoming more negative, then changing towards positive. The EEMI from the phason mode in ${\rm YMn}_{6}{\rm Sn}_{6}$ is anticipated to show the resonance-type current-density dependence; it may stay negative when the pinning frequency (${\omega}_{\rm pin}$) is larger than the ac current frequency (${\omega}_{\rm obs}$), while changing to positive for the ${\omega}_{\rm pin}$ reduced below ${\omega}_{\rm obs}$ by the increased current density due to the phason depinning transition~\cite{fref16, fref19}. The behavior of the inductance for the Tb 0 $\%$ compound at 300 K and 1.5 T may reflect such a crossover at ${\omega}_{\rm pin} \sim {\omega}_{\rm obs}$ around ${j}_{0} \sim\ $4$\times {10}^{4}$ A/${\rm cm}^{2}$, while the EEMI signal therein may include the contribution from the current-induced depinning transition of the H-AF domain wall excitations as described above. As opposed to the complex current density dependence of the phason-mode inductance, the monotonous current-density dependence is always observed for both the positive inductance of the AF phase (Tb 0 $\%$, 300 K, 3.5 T), and the high-temperature magnetic phase boundary region (Tb 7 $\%$). This is in accord with the scenario in which the spin fluctuation-enhanced inductance is in line with the tilting mode mechanism irrelevant to the depinning current density.

\begin{figure*}
\begin{center}
\includegraphics*[width=7in,keepaspectratio=true]{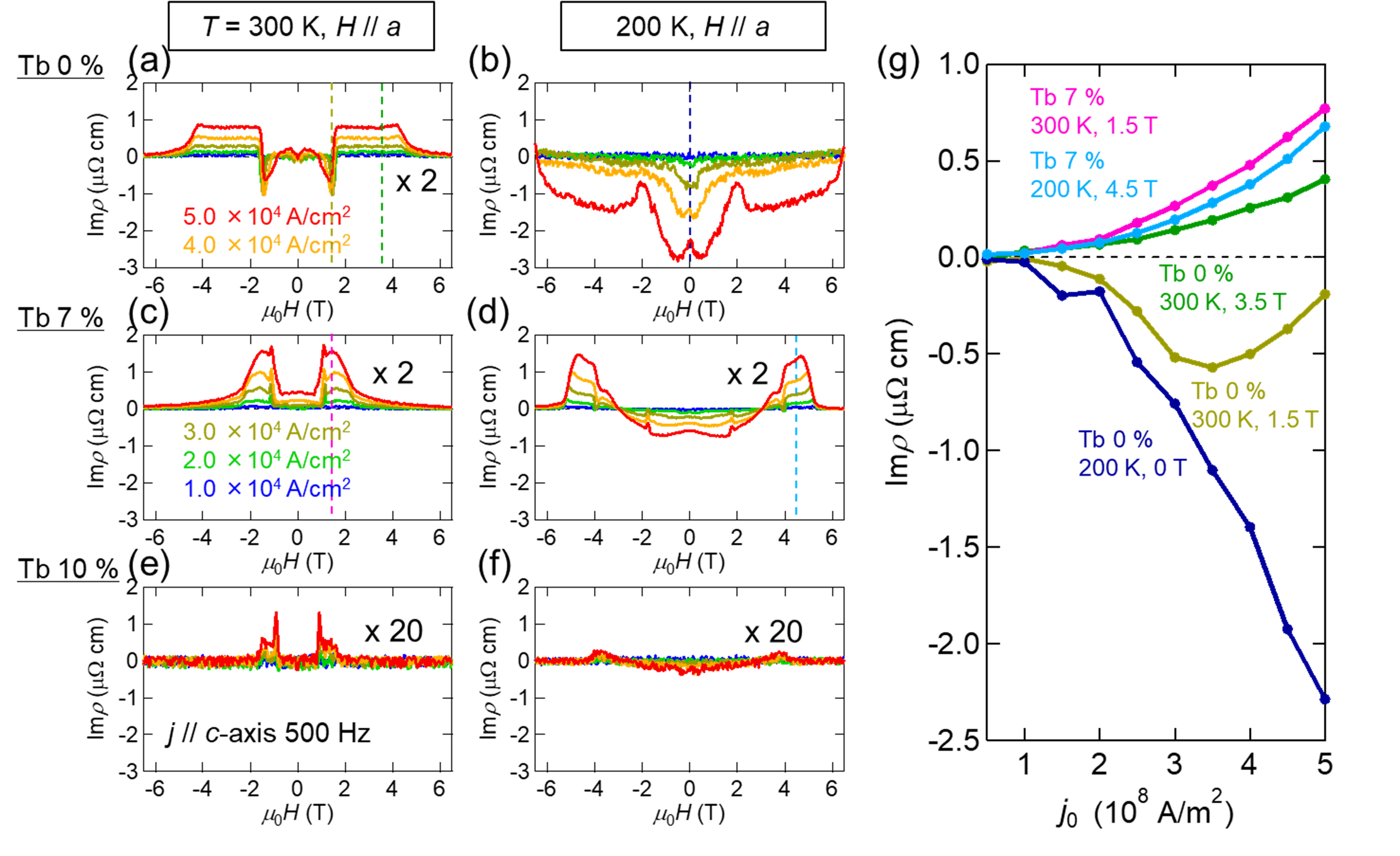}
\caption{(a)-(f) Current-density (${j}_{0}$) dependence of the imaginary part of complex resistivity Im $\rho$ measured under $H \parallel a$. g) Current density (${j}_{0}$) dependence of Im $\rho$ at some selected fixed magnetic fields as denoted by vertical dashed lines in (a)-(d).}
\end{center}
\end{figure*}

\section{VI. Conclusion}
We have investigated the impact of chemical doping on both the collinear/noncollinear magnetic structures and the emergent electromagnetic induction near room temperature in ${\rm YMn}_{6}{\rm Sn}_{6}$. The SANS experiments on Tb-doped ${\rm Y}_{1-x}{\rm Tb}_{x}{\rm Mn}_{6}{\rm Sn}_{6}$ crystals have clarified that the incommensurate spin-helix period increases from ${\it \lambda}$ = 2.4 nm (300 K, 0 T) at $x$ = 0 to ${\it \lambda}$ = 3.7 nm (300 K, 0 T) for $x$ = 0.1, and the collinear double-antiferromagnetic (AF) state existing near the high-temperature magnetic phase boundary for $x$ = 0 is eliminated $x$ $\geq$ 0.07. The latter effect is likely due to the introduction of pinning centers via Tb doping. Our systematic investigation of the emergent electromagnetic inductance (EEMI) in ${\rm (Y, Tb)Mn}_{6}{\rm Sn}_{6}$ crystals has clarified the following features; (1) The negative EEMI observed in the low-temperature helix phase is greatly suppressed upon Tb doping, and (2) the positive inductance at relatively high temperatures near the magnetic phase boundary mostly survives against Tb doping. As for the feature (1), for example, the negative inductance value at 200 K is conspicuously reduced to 20 $\%$ and $<$ 1 $\%$ upon Tb doping of 7 $\%$ and 10 $\%$, respectively. This reduction of the EEMI upon Tb doping is too large to be ascribed simply to the change of magnetic period ${\it \lambda}$ of the incommensurate helix; the EEMI would be in proportion to ${\it \lambda}^{-1}$, if other characteristics were the same. The negative EEMI is necessarily caused by the phason mode dynamics driven by the ac current. Therefore, the drastic change in negative EEMI can be ascribed to an impurity (Tb) induced pinning effect on the dynamics of the spin helix. As for the feature (2), the positive inductance is anticipated to stem generally from the tilting mode of the helix~\cite{fref16, fref19}. We find that in the present case, it can appear around the high temperature phase boundary regardless of whether the dominant magnetic order is commensurate (collinear) AF or incommensurate (noncollinear) helix. This fact leads us to conclude that this positive inductance is also derived from the spin noncollinearity induced via magnetic fluctuations near the phase boundary; the mechanism can be regarded as an extension of the tilting-mode one. 
The present investigation on the chemically doped helimagnets allows us to propose useful disciplines of materials design for the high-temperature, i.e., around/beyond room temperature, EEMI. To promote the negative EEMI, in addition to the short-period (a few nm) helix and high conductivity of the material, crystals of high-purity, low levels of imperfection, and low magnetic anisotropy will be favorable to enhance the phason mode dynamics exerted by ac current. Such a negative EEMI can potentially be turned into the large positive EEMI with increasing the exciting current density through the current-induced phason depinning transition. On the other hand, to target the positive emergent inductance, the introduction of pinning centers like those due to impurity doping, and/or the increase of the thermal spin fluctuations, will suppress the phason mode but keep the tilting mode relatively intact. This situation thus favors the tilting mode-induced positive inductance component that is otherwise cancelled or overwhelmed by a competing negative component. The present observation of the positive inductance under the thermal agitation provides a hint for expanding the range of emergent inductor candidate materials to the collinear-spin magnetic materials.

\section{Acknowledgements}
This work was supported by Core Research for Evolutional Science and Technology (CREST), Japan Science and Technology Agency (JST) (Grant No. JPMJCR1874 and No. JPMJCR16F1), Fusion Oriented Research for Disruptive Science and Technology (FOREST), Japan Science and Technology Agency (JST) (Grant No. JPMJFR2038), the Japan Society for the Promotion of Science (JSPS) KAKENHI (Grant No. JP20H01859, JP20H05155 and No. JP21J11830), the Swiss National Science Foundation (SNSF) Sinergia network “NanoSkyrmionics” (Grant No. CRSII5\_171003), the SNSF Project No. 200021\_188707 and an ETH Z\"{u}rich Research Partnership Grant RPG\_072021\_07. This work is based partly on experiments performed at the Swiss spallation neutron source SINQ, Paul Scherrer Institute, Villigen, Switzerland.

\newpage
\onecolumngrid
\setcounter{equation}{0}
\setcounter{figure}{0}
\setcounter{table}{0}
\setcounter{page}{1}
\makeatletter
\renewcommand{\theequation}{S\arabic{equation}}
\renewcommand{\thefigure}{S\arabic{figure}}
\renewcommand{\bibnumfmt}[1]{[S#1]}

\renewcommand{\citenumfont}[1]{S#1}

\clearpage

\maketitle

\section*{Supplemental Material}

\section*{Magnetic properties of ${\bf (Y, Tb)Mn}_{6}{\bf Sn}_{6}$}

In Figs.~S1-S3, we show the magnetization curves of ${\rm (Y, Tb)Mn}_{6}{\rm Sn}_{6}$. Magnetization measurement was performed by using Quantum Design PPMS-14 T with ACMS option. While any hysteresis does not appear in Tb 0 $\%$ and 7 $\%$ case, a hystereic and additional transition below 100 K appears in Tb 10 $\%$ case.

\begin{figure*}[h]
\begin{center}
\includegraphics*[width=7in,keepaspectratio=true]{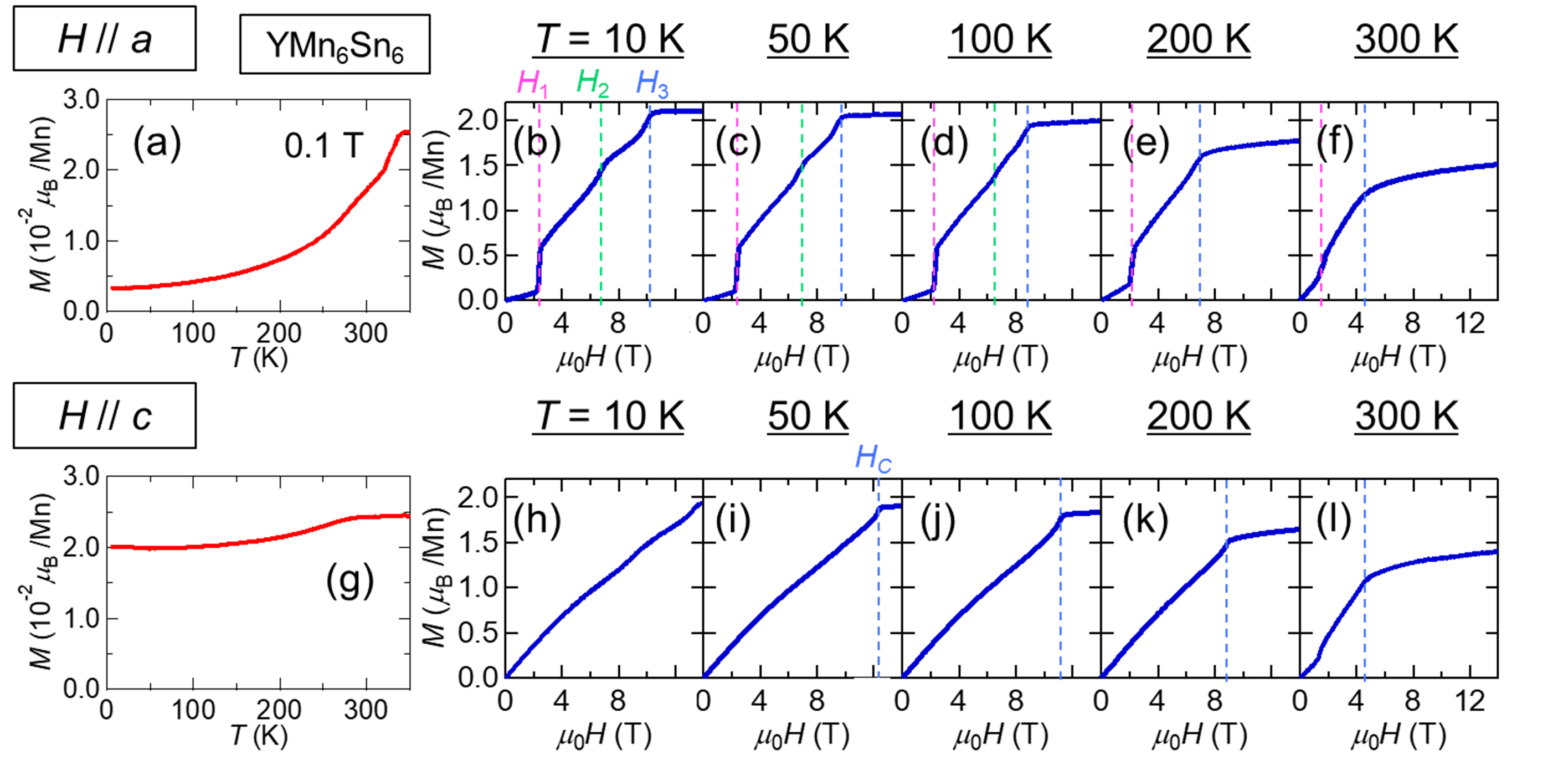}
\caption{Magnetization curves of ${\rm YMn}_{6}{\rm Sn}_{6}$. (a) Temperature dependence of magnetization $M$ at 0.1 T along $a$-axis. The $M$-$T$ curves under field cooling (FC) and zero field cooling (ZFC) processes remain the same. (b)-(f) Magnetic field dependence of magnetization of ${\rm YMn}_{6}{\rm Sn}_{6}$ at 0.1 T along $a$-axis. (g) Temperature dependence of magnetization at 0.1 T along $c$-axis. The $M$-$T$ curves under FC and ZFC processes remain the same. (h)-(l) Magnetic field dependence of magnetization of ${\rm YMn}_{6}{\rm Sn}_{6}$ at 0.1 T along $c$-axis. In all temperatures, no hysteresis is observed.}
\end{center}
\end{figure*}

\begin{figure*}[h]
\begin{center}
\includegraphics*[width=7in,keepaspectratio=true]{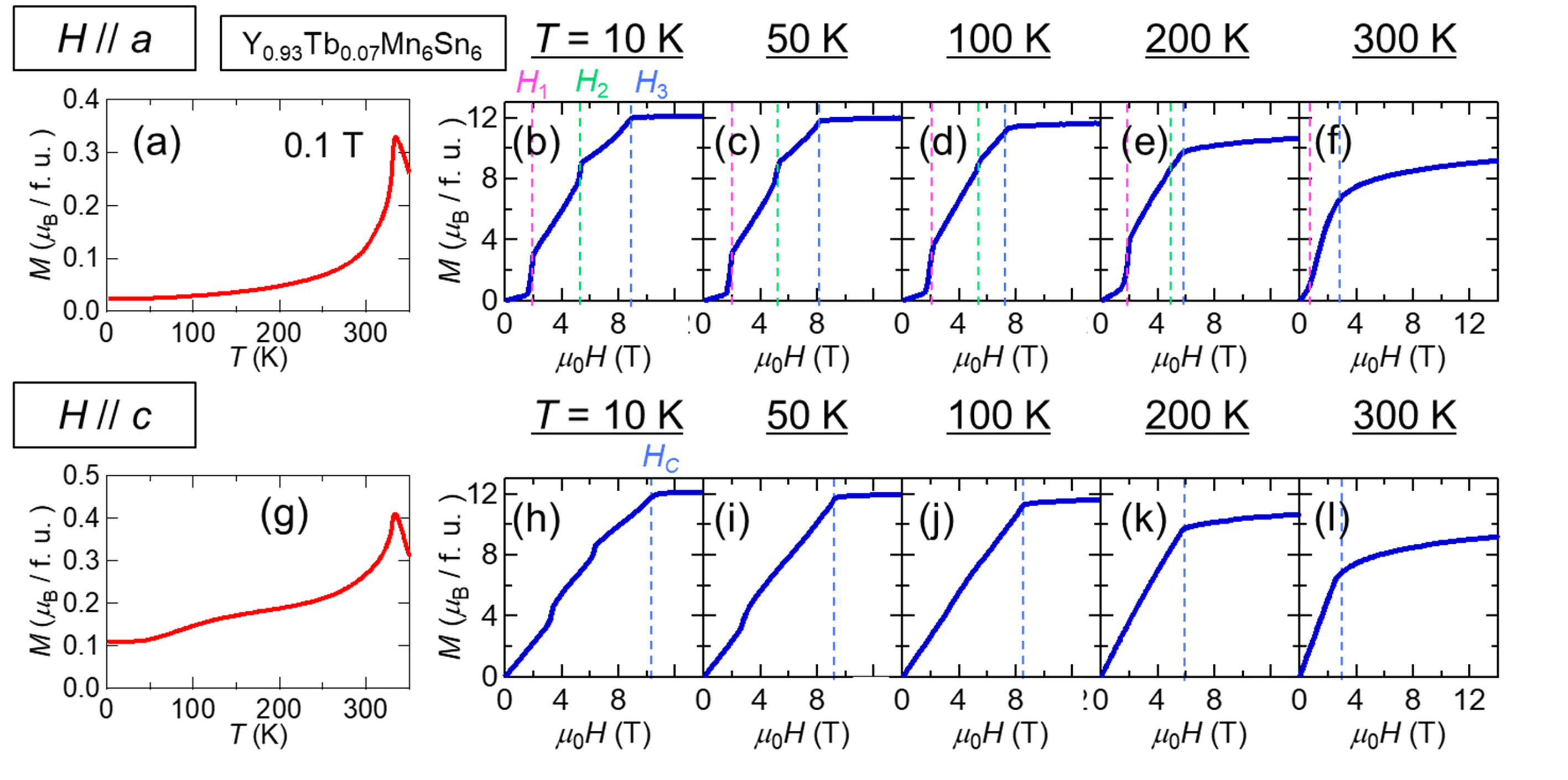}
\caption{Magnetization curves of ${\rm Y}_{0.93}{\rm Tb}_{0.07}{\rm Mn}_{6}{\rm Sn}_{6}$. (a) Temperature dependence of magnetization M at 0.1 T along $a$-axis. The $M$-$T$ curves under field cooling (FC) and zero field cooling (ZFC) processes remain the same. (b)-(f) Magnetic field dependence of magnetization of ${\rm Y}_{0.93}{\rm Tb}_{0.07}{\rm Mn}_{6}{\rm Sn}_{6}$ at 0.1 T along $a$-axis. (g) Temperature dependence of magnetization at 0.1 T along $c$-axis. The $M$-$T$ curves under FC and ZFC processes remain the same. (h)-(l) Magnetic field dependence of magnetization of ${\rm YMn}_{6}{\rm Sn}_{6}$ at 0.1 T along $c$-axis. In all temperatures, no hysteresis is observed.}
\end{center}
\end{figure*}

\begin{figure*}[h]
\begin{center}
\includegraphics*[width=7in,keepaspectratio=true]{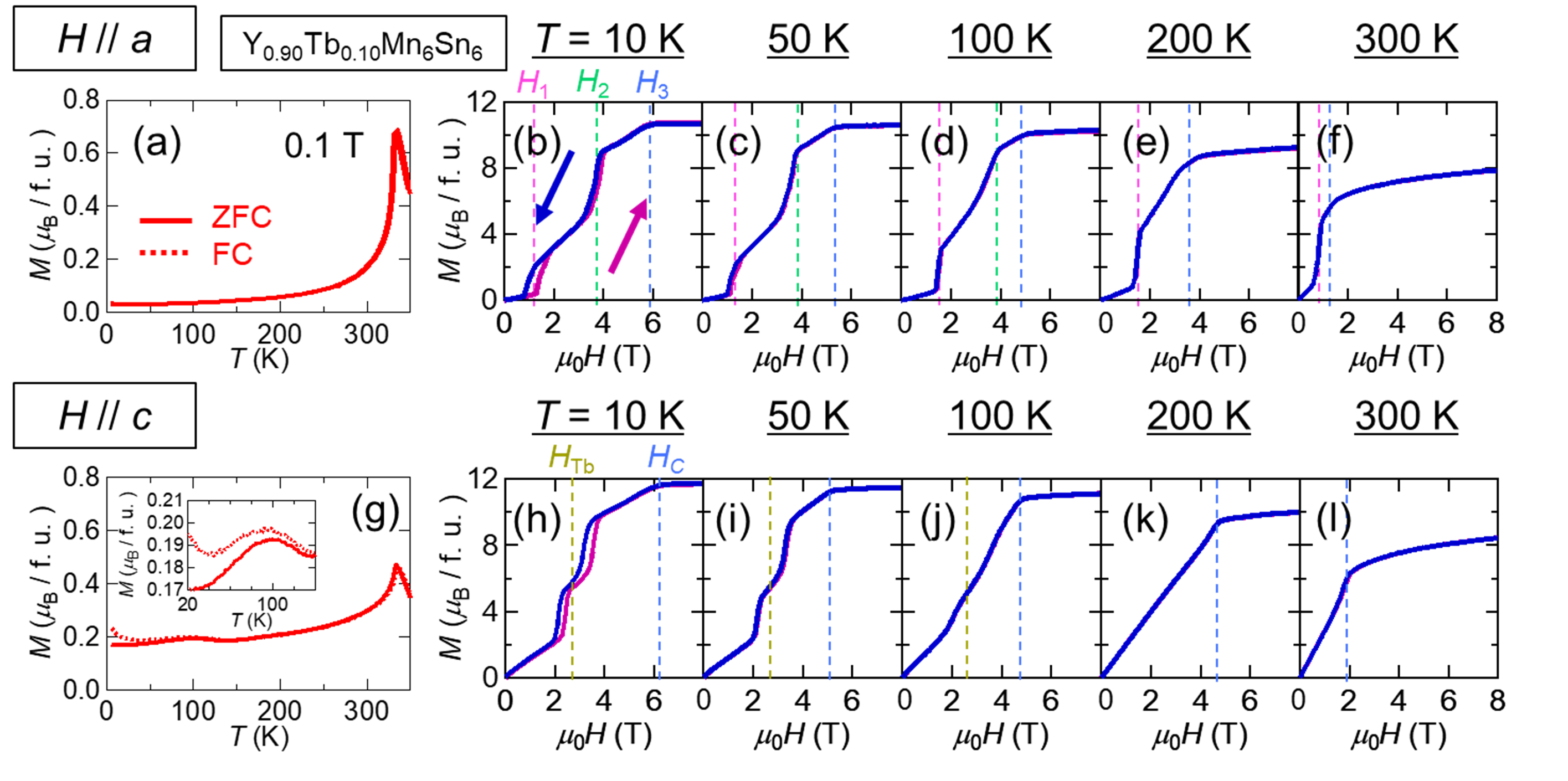}
\caption{Magnetization curves of ${\rm Y}_{0.90}{\rm Tb}_{0.10}{\rm Mn}_{6}{\rm Sn}_{6}$. (a) Temperature dependence of magnetization M at 0.1 T along $a$-axis. The dotted and solid lines correspond to $M$-$T$ curves of field cooling (FC) and zero field cooling (ZFC) processes, respectively. (b)-(f) Magnetic field dependence of magnetization of ${\rm YMn}_{6}{\rm Sn}_{6}$ at 0.1 T along $a$-axis. Blue and purple curves correspond to $M$-$H$ curves in the magnetic field descending and increasing processes, respectively. (g) Temperature dependence of magnetization at 0.1 T along $c$-axis. The dotted and solid lines correspond to M-T curves of FC and ZFC processes, respectively. The inset indicates the magnified view around 100 K. (h)-(l) Magnetic field dependence of magnetization of ${\rm YMn}_{6}{\rm Sn}_{6}$ at 0.1 T along $c$-axis. Blue and purple curves correspond to $M$-$H$ curves in the magnetic field descending and increasing processes, respectively.}
\end{center}
\end{figure*}

\clearpage

\section*{Nonmonotonic spiral structure of ${\bf YMn}_{6}{\bf Sn}_{6}$}

${\rm YMn}_{6}{\rm Sn}_{6}$ has nonmonotonically-winding spiral structures. In SANS measurement data, the intensity ratio of 2$\pi$/$c - {q}_{\rm IC}$ and ${q}_{\rm IC}$ exemplifies this feature. In the case of monotonically-winding spiral structure, the angle of Mn spins between nearest neighbor sites (${\alpha}$) become the half of that between second nearest neighbor sites (${\beta}$). The definition of ${\alpha}$ and ${\beta}$ are illustrated in Figs.~S4(a).
We can calculate the tilting angle from SANS intensity of 2${\pi}$/$c - {q}_{\rm IC}$ [$I$(2$\pi$/$c - {q}_{\rm IC}$)] and that of ${q}_{\rm IC}$ [$I$($q$)] with the assumption that ${\alpha}$ and ${\beta}$ are constant even under $H \parallel a$, such as
\begin{eqnarray*}
\alpha=&&\pm{\rm cos}^{-1}\left(\frac{\pm A}{B}\right)\\
A=&&\sqrt{\frac{I\left(\frac{2{\pi}}{c}-q\right)}{I\left(q\right)}}{\rm sin}\left(qc\left(\frac{1}{2}-z\right)\right)-{\rm sin}\left(qc\left(\frac{1}{2}-z\right)\right){\rm cos}\left(2{\pi}z\right)-{\rm cos}\left(qc\left(\frac{1}{2}-z\right)\right){\rm sin}\left(2{\pi}z\right)\\
B=&&\frac{I\left(\frac{2{\pi}}{c}-q\right)}{I\left(q\right)}{\rm sin}^{2}\left(qc\left(\frac{1}{2}-z\right)\right)
+\frac{I\left(\frac{2{\pi}}{c}-q\right)}{I\left(q\right)}{\rm cos}^{2}\left(qc\left(\frac{1}{2}-z\right)\right)\\
&-&\sqrt{\frac{I\left(\frac{2{\pi}}{c}-q\right)}{I\left(q\right)}}{\rm cos}^{2}\left(qc\left(\frac{1}{2}-z\right)\right){\rm cos}\left(2{\pi}z\right)\\
&-&\sqrt{\frac{I\left(\frac{2{\pi}}{c}-q\right)}{I\left(q\right)}}{\rm sin}^{2}\left(qc\left(\frac{1}{2}-z\right)\right){\rm cos}\left(2{\pi}z\right)\\
&+&{\rm sin}^{2}\left(qc\left(\frac{1}{2}-z\right)\right){\rm sin}^{2}\left(2{\pi}z\right)
+{\rm cos}^{2}\left(qc\left(\frac{1}{2}-z\right)\right){\rm cos}^{2}\left(2{\pi}z\right)\\
&+&{\rm cos}^{2}\left(qc\left(\frac{1}{2}-z\right)\right){\rm sin}^{2}\left(2{\pi}z\right)
+{\rm sin}^{2}\left(qc\left(\frac{1}{2}-z\right)\right){\rm cos}^{2}\left(2{\pi}z\right)
\end{eqnarray*}

Here, $q$, $c$ and $z$ are the length of propagation vector ${\bm q}_{\rm IC}$, the unit cell length along $c$-axis (0.898 \AA), and the relative position of Mn1 atom (0.2496) along $c$-axis in the unit of $c$/2${\pi}$, respectively. ${\beta}$ is proportional to $q$ [${\beta}$(\(^\circ\)) = 257.4 $\times$ $q$ (${\rm \AA}^{-1}$)]. Figs.~S4(b)-(d) show the temperature and magnetic field dependence of ${\alpha}$/${\beta}$ and Figs.~S4(e)-(f) show the temperature and magnetic field dependence of $q$. The ratio ${\alpha}$/${\beta}$ ranges from 0.37 to 0.42.

\begin{figure*}[h]
\begin{center}
\includegraphics*[width=5in,keepaspectratio=true]{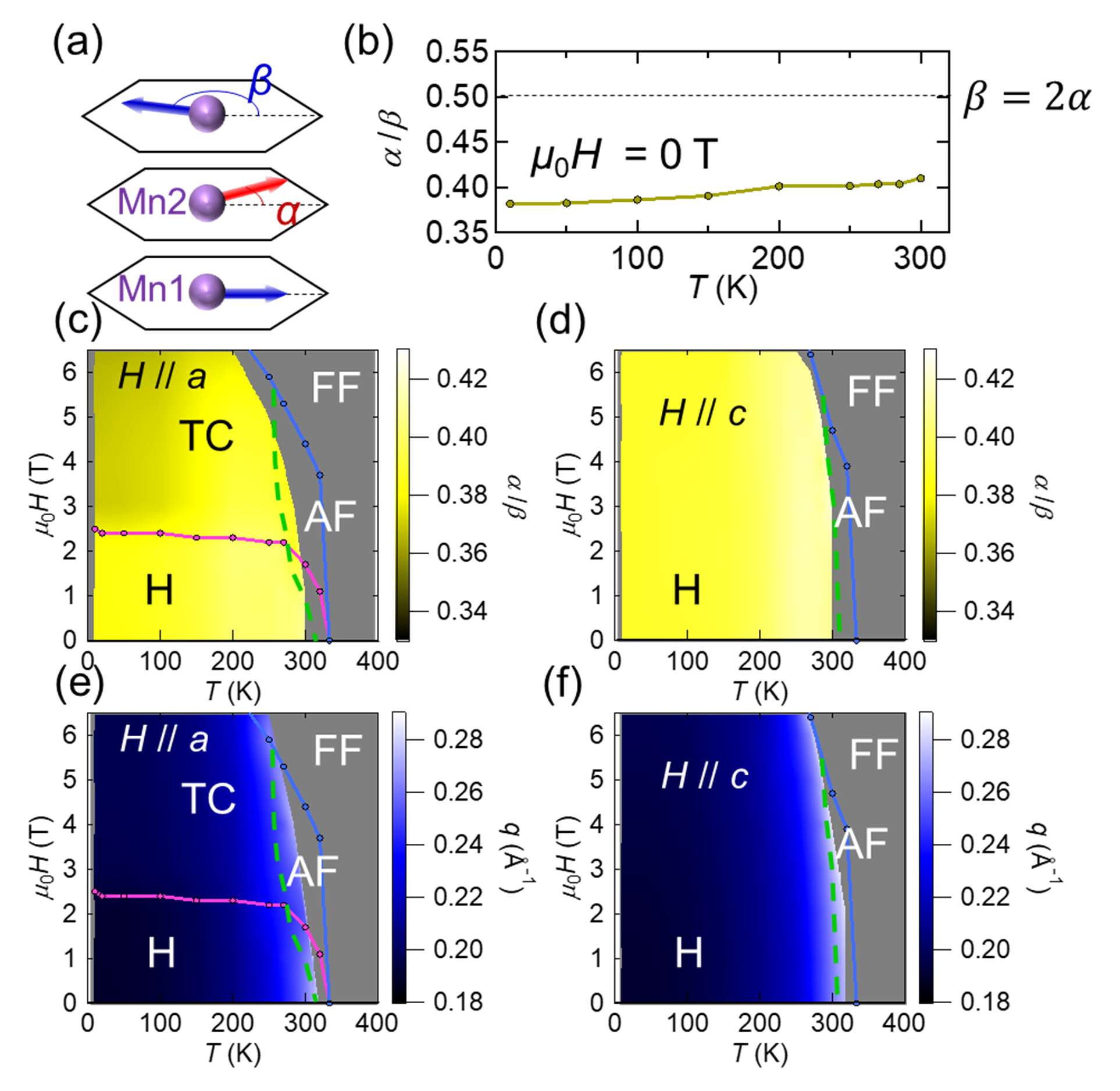}
\caption{(a) Schematic illustration showing the interplane angles of spins in a nonmonotonically-winding spiral structure. (b) Temperature dependence of ${\alpha}$/${\beta}$ in ${\rm YMn}_{6}{\rm Sn}_{6}$. In the monotonically-winding spiral structure, it becomes 0.50 (dashed line). (c)-(d) Color maps of ${\alpha}$/${\beta}$ on the magnetic phase diagrams. (e)-(f) Color maps of $q$, length of propagation vector on the magnetic phase diagrams. $q$ and ${\beta}$ are connected by the relation that ${\beta}$(\(^\circ\)) = 257.4 $\times$ $q$ (${\rm \AA}^{-1}$).}
\end{center}
\end{figure*}

\clearpage

\section*{SANS results of ${\bf Y}_{0.90}{\bf Tb}_{0.10}{\bf Mn}_{6}{\bf Sn}_{6}$ (${\bf Tb}$ 10 $\%$) under magnetic fields along the ${\bm C}$-axis}

SANS results for ${\rm Y}_{0.90}{\rm Tb}_{0.10}{\rm Mn}_{6}{\rm Sn}_{6}$ (Tb 10 $\%$) crystal are shown in Figs.~S5. The incommensurate helix at zero field appears to be continuously transformed to the longitudinal cone state and finally turned in to the forced ferromagnetic (FF) state, as seen in the reduction of the diffraction intensity with the magnetic field applied along the $c$-axis, while the magnnetic modulation ${q}_{\rm IC}$ is kept almost unchanged.

\begin{figure*}[h]
\begin{center}
\includegraphics*[width=4in,keepaspectratio=true]{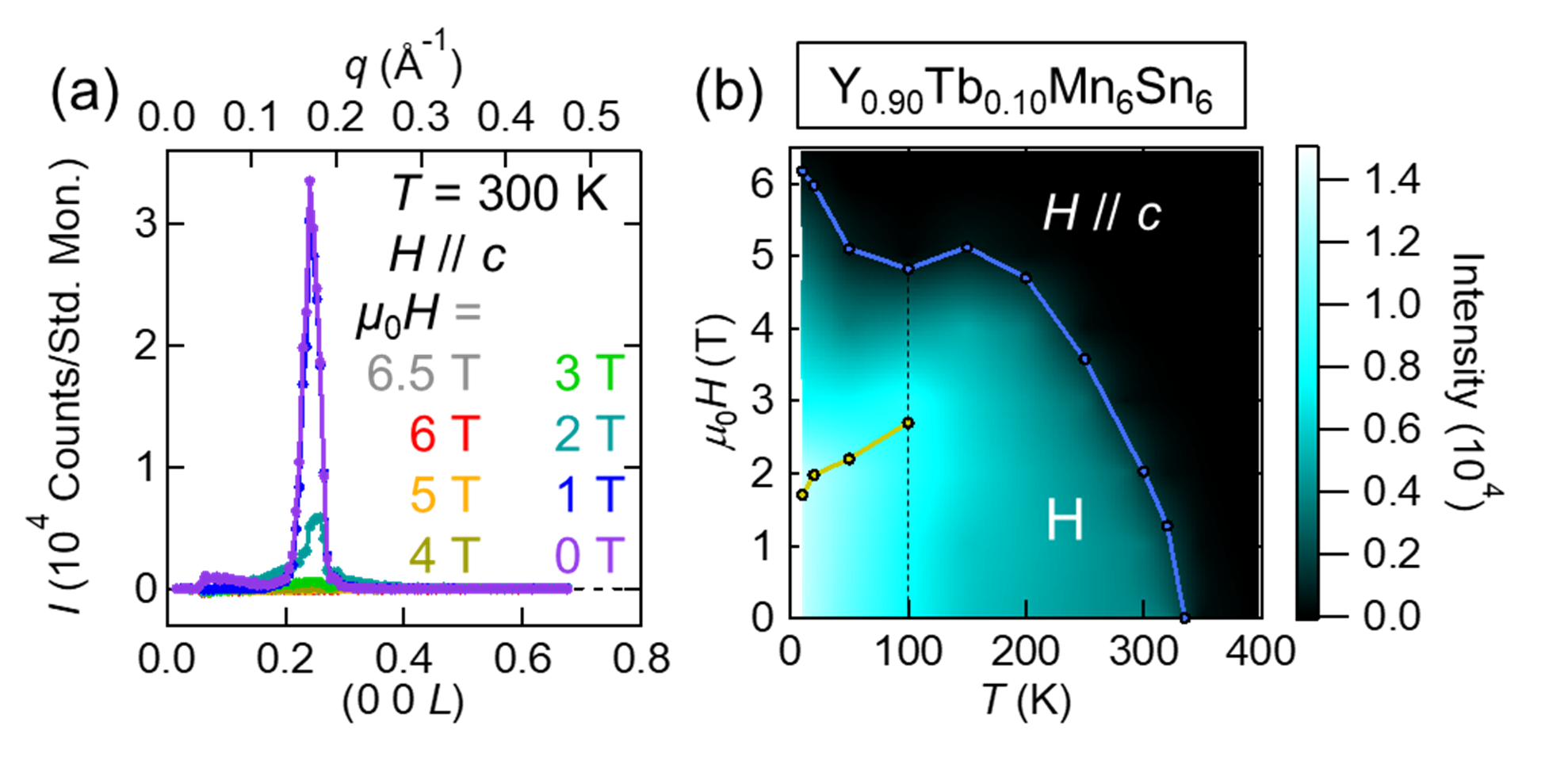}
\caption{SANS results of ${\rm Y}_{0.90}{\rm Tb}_{0.10}{\rm Mn}_{6}{\rm Sn}_{6}$ (Tb 10 $\%$) under various magnetic fields applied along the $c$-axis. (a) The plot of SANS intensity vs. $q$ at 300K.  (b) The SANS intensity contour maps on the magnetic phase diagram.}
\end{center}
\end{figure*}

\end{document}